\documentclass[aps,twocolumn,10pt]{revtex4}  
\usepackage{graphicx}
\usepackage[dvips]{color}
\input{epsf.tex}
\newcommand{\bibi}{\bibitem}                                                  
\newcommand{\etal}{\it {et al.}}                                             

\newcommand{\half}{\frac {1}{2}}                                             
                                             
\newcommand{\beq}{\begin{equation}}                                           
\newcommand{\eeq}{\end{equation}\noindent}                                  
\newcommand{\beqr}{\begin{eqnarray}}                                          
\newcommand{\eeqr}{\end{eqnarray}\noindent}                                   
\newcommand{\vQ}{\bf Q}                                                      
\newcommand{\vcr}{{\bf r}}                                                     
\newcommand{\vk}{{\bf k}}                                                     
                                                     
\newcommand{\vq}{{\bf q}}

\newcommand{\cd}{c^{\dag}} 
\newcommand{\bd}{b^{\dag}} 
\newcommand{\hd}{h^{\dag}}

\newcommand{\noin}{\noindent}                                                 
\begin{document} 
%\twocolumn[\hsize\textwidth\columnwidth\hsize\csname@twocolumnfalse\endcsname
%\onecolumn
\title{A Consistent Theory of Underdoped Cuprates: Evolution of the RVB State From Half Filling}
\author{Sanjoy K. Sarker and Timothy Lovorn\\                                                   
Department of Physics and Astronomy \\                                        
The University of Alabama, Tuscaloosa, AL 35487 \\}                           
%\date{}
%\begin{abstract}                                                                       
%\maketitle    
%\widetext
\begin{abstract}

Using continuity, we derive a renormalized Hamiltonian from the 
parent $t$-$J$ model to describe the properties of underdoped cuprates. 
The theory is constrained to agree with the behavior at half filling, 
which is well described by the Arovas-Auerbach valence-bond state in 
which bosonic spinons are paired into singlets. Spinon states evolve 
continuously into the doped region preserving their symmetry. We assume that 
moving holes rapidly destroy magnetic order, which leads to a gap in the 
spinon spectrum and strongly renormalizes the theory. 
The spin gap leads to two new types hopping terms for renormalized holes. 
In one, a fermionic holon hops within the same
sublattice accompanied by a singlet backflow, 
giving rise to a non-Fermi liquid normal state with novel properties. 
Spinon singlets condense below a pseudogap temperature $T^*$
($<$ spin gap temperature $T^0$), which allows holons to propagate coherently, 
forming a spinless Fermi liquid, but without an observable holon Fermi surface. 
Above $T^*$, holons are localized. 
This is the so-called strange metal phase, which is actually  
a new type of insulator. In the second term a pair of holons belonging to 
opposite sublattices hop, accompanied by a singlet backflow. 
In the presence of the singlet condensate holon pairs condense, 
leading to $d$-wave superconductivity; the symmetry is primarily determined by the 
symmetry of the valanece-bond state at half filling. The metal and the superconductor
preserve the two-sublattice character of the valence-bond state.
A careful examination of the nmr, tunneling and 
transport data shows that the predictions of the theory is consistent with experimental results. 
Remarkably, the existence of the spin gap provides a natural explanation for the 
phenomenon of two-dimensionality of the normal state in the presence of interplane 
hopping.  The marked asymmetry between hole-doped and electron-doped cuprates
is also easily explained.

\end{abstract} 
\maketitle                                                               
\vspace{0.5 in}                                                              

%\pagebreak                                                                    

%\narrowtext
%\twocolumn

\noin{\bf{I. Introduction}}
\\

The origin of high-temperature superconductivity in doped cuprates \cite{bed} is 
believed to be closely linked with the unusual behavior of the normal state,
which is not a Fermi liquid. Furthermore, in the underdoped region a pseudogap 
appears below a temperature $T^*$, much above the superconducting temperature 
$T_c$ \cite{tim}. Even before the discovery of the pseudogap phase, Anderson
argued that the non-Fermi liquid behavior is due to the two-dimensional (2d) 
nature of the normal state \cite{and}, and its proximity to the undoped phase, 
which is a Mott insulator, or equivalently, a quantum antiferromagnet. 
The 2d behavior is unexpected, and its origin is not understood. Anderson also
proposed the resonating-valence-bond (RVB) state, in which spins are paired 
into singlets within the insulator \cite{and}. Upon doping with holes, this state would evolve 
continuously into the doped region where spinless holons, which carry charge, would propagate 
coherently and create a metal. The elementary excitations are 
holons and spin-1/2 spinons. The physical electron is a composite particle. The 
expectation is that the apparent spin-charge separation would lead to non-Fermi liquid 
behavior, and further that the spin singlets would acquire charge via some 
interaction with holons and become superconducting. 

Theoretical studies are usually based on the large-$U$ Hubbard model, or, equivalently, 
the $t$-$J$ model on a square lattice, where $J = 4t^2/U$ is the exchange coupling
between spins, and $t$ describes nearest-neighbor hopping of electrons such that
no site is doubly occupied. For cuprates, $t/J \sim 3 - 4$. The constraint of no
double occupancy corresponds to a $U(1)$ gauge symmetry \cite{bas1}. Much of the 
work has been devoted to deriving an effective low-energy theory of 
propagating spinons and holons (or pairs), coupled to a gauge field. In principle,
such a task should not be too diffcult since one can invoke continuity.
The effective action would depend on the symmetries of the underlying 
vacuum state, i.e., a renormalized version of the RVB state of paired spinons,
which is presumed to describe the insulator at half filling. After more than two 
decades of intensive work, a successful theory has not been found. 

This failure is surprising since, using a mean-field approximation,
Arovas and Auerbach \cite{aro} have shown that the behavior at half filling is indeed 
reasonably well described by an RVB state in which singlets formed by pairing bosonic 
spinons (Schwinger bosons) condense below a temperature $T_{RVB}$. (In this paper we
use the term RVB to describe any valence-bond state). One complication is that 
a fraction of spinons remains unpaired, and condenses separately to give rise to 
long-range antiferromagnetic order. This is not a defect of the theory since it 
correctly describes the ground state which is known to be ordered \cite{sar1}.  
In other words, the system is in a mixed phase of an RVB state 
and a N\'eel state, with $T_{RVB} > T_{AF}$ (in $d =2$, $T_{AF} = 0$). In cuprates
the AF insulating state exists up to a hole density $x$ of about $.05$, 
beyond which there is a transition to a pseudogap metal (and a $d$-wave superconductor), 
with no long-range magnetic order of any type. However, a seemingly straightforward 
extention of the Arovas-Auerbach MF theory (with fermionic holons)
to the doped region leads to a metallic state with spiral magnetic order \cite{sar2}. 
This failure has lead to a virtual abandonment of the approach based on the 
Arovas-Auerbach RVB state and bosonic spinons \cite{weng}.

Instead, RVB theories based on the slave-boson representation (spinons are
fermions, holons are bosons) have been used widely \cite {bas,kot,ichi} 
(for a review, see Lee {\etal} \cite{lee1}). But these
do not work at half-filling, and to our knowledge, there are no 
experimental signatures of a transition to a different RVB state accompanied
by statistical transmutation upon doping. There are serious doubts as to their
efficacy in describing the physics of underdoped cuprates. Indeed the 
failure to connect with the physics at half-filling is one of the central problems 
in high-$T_c$ theory. 

We have solved this problem by deriving a renormalized theory for small $x$ which 
is consistent with the physics at half filling, but does not have the problems of 
spiral instability. The key point is that, for the $t$-$J$ model, the 
spiral state has been shown to be unstable \cite{hu}.  
We therefore assume that (1) the spin states evolve continuously from half filling, 
preserving their symmetry and particle statistics, and (2) moving holes rapidly destroy 
long-range magnetic order at $T = 0$ beyond some small critical concentration $x_c$, 
strongly renormalizing the theory in the process; but the VB state survives 
up to a temperature $T_{RVB}(x)$. These are reasonable assumptions, 
and formed the basis of early gauge theories based on the
Schwinger-boson representation \cite{weig,lee3}. But in those theories it was {\em assumed} 
that the initial renormalization leads to a sublattice-preserving $t^{\prime}$-$J$ model, 
where $t^{\prime}$ is an effective next-nearest neighbor hopping parameter.  

The actual renormalized Hamiltonian is found to be quite different. A brief account 
of the theory focusing primarily on the pseudogap phase and some of the results 
have been reported earlier \cite{sar9}. Here we provide a detailed description, 
and show that it not only accounts for the main features of the phase diagram, 
but predicts novel properties of the underlying states that require a major 
revision of our current understanding of the cuprate phenomenology. Additionally,
the theory provides a natural explanation for the two-dimensionality of the normal
state, as well as for the marked asymmetry between hole-doped and electron-doped 
cuprates. 

The continuity requirement ensures that there are exactly the 
same three spinon states in the doped region as at half filling, 
and holon motion do not introduce any new order 
parameters {\em for the spinons}. Destruction of magnetic order
automatically leads to a gap for bosonic spinons in the other two states: 
the RVB state, and the nonordered state \cite{aro,read}. 
The spin gap - not to be confused with 
the pseudogap - is the distinctive feature in the theory. As described in 
section IV, the gap allows us to renormalize away nearest-neighbor 
hopping ($t$), and use perturbation theory to obtain a minimal 
Hamiltonian involving sublattice-preserving hopping of renormalized holes, 
accompanied by singlet backflows. Two types of new hopping terms are generated:
in one a single holon hops within the 
same sublattice, and in the other a pair of holons hop together.

We analyze the Hamiltonian in section V, showing that the one-hole term
gives rise to exactly two {\lq normal'} states, as found in cuprates, whose
symmetries are determined by the underlying spinon states (at half filiing)
-- no new order parameter is introduced, either in the spin or the charge sector.
In the absence of singlet condensation (i.e., above $T_{RVB}(x)$), holons are localized.  
We identify this state with the so-called strange metal phase, 
except that it is not a metal, but a new kind of insulator, 
which is an important prediction. Coupling to the RVB condensate allows holons to 
hop coherently within the same sublattice, and create a spinless Fermi liquid of 
concentration $x$ below $T_{RVB}(x)$, which we identify with the pseudogap 
temperature $T^*$. However, gauge invariance ensures that there is no observable 
(small) Fermi surface, which is another prediction.  

As shown in section VI, in the presence of a RVB condensate  
the pair-hopping term allows holon pairs to Bose condense,  
giving rise to $d$-wave superconductivity. The symmetry is largely determined 
by that of the underlying RVB state 
known from half filling. As discussed in section II, the valence-bond state has 
a two-sublattice property of its own: on average, singlets connect spins on opposite 
sublattices. This property is predicted to be preserved in the pseudogap metal and 
superconducting phases.  

The absence of any new order (other than superconductivity) imposes constraints
on the number and type of phases; the resulting phase diagram is
in qualitative agreement with the experimental one. 
The constraints allow us to extract many universal
features from the structure of the Hamiltonians in various phases, without detailed 
calculations or fine tuning. For example, the spin gap by itself rules out a 
conventional Fermi liquid state for small $x$. 
In section VII we carefully examine the existing experimental results, 
and find strong support for the predictions of our theory. In particular, results from nmr, 
transport, optical conductivity and tunneling measurements, taken together, provide strong
evidence for a spin gap, a nonmetallic phase above $T^*$, and 
a spinless Fermi liquid without a Fermi surface below.  
We will point out that these results can not be reconciled with either a Fermi
liquid, or earlier RVB theories.

In section VIII we discuss the fundamental issue of the two-dimensionality of 
the normal state, which is unexpected since the cuprates are 3d - though layered - 
materials, and localization within a plane would cost too much kinetic energy. 
Indeed, all the other nearby states: the antiferromagnet in the undoped phase, 
the superconducting state, and the Fermi liquid state that appear at large doping 
exhibit 3d behavior. Hence theories that are based on the 2d model are suspect if 
they fail to exhibit 
2d confinement when hopping in the perpedicular direction is turned on. Indeed,
such a test will probably
invalidate most of the existing theories.  We show that, in our case, the spin gap provides a 
protective mechanism against delocalization of holons in the $z$ direction, which
is not obvious since holons are spinless. However, the holon pairs can
hop coherently, which is essential for the observed 3d superconductivity. 
We solve the 3d problem and show that condensation energy is enhanced due 
to 3d coupling. 

Another issue that is not completely understood is the significant
difference between the hole-doped and electron-doped cuprates. In section IX 
we will argue that the difference is due to direct
intra-sublattice hopping $t^{\prime}$, which breaks electron-hole symmetry in the 
Hubbard model. Normally the difference is not significant, we will present results to 
show that in our case it is substantially enhanced becuse the nearest-neighbor hopping 
($t$) is renormalized away. The issue is subtle and depends on the symmetry of 
the underlying spinon states. 

Our renormalized Hamiltonian belongs to the class of short-range RVB models considered by
Kivelson, Rokhsar and Sethna \cite{kiv,rokh}. However, because of the consistency
requirements the Hamiltonian differs significantly from those proposed earlier.
The present theory is based on a paper published much earlier, in which
a self-consistent perturbation expansion in powers of hopping was used
to study the destruction of long-range magnetic order and the occurrence of 
non-Fermi liquid behavior \cite{sar3}. The intra-plane pair hopping mechanism 
for superconductivity in the present theory was also discovered there. An effective
Hamiltonian for renormalized holons was derived in ref. \cite{sar5}, but with an (assumed) 
wrong (spiral) normal state. In section III we review these and earlier works by other 
authors that are relevant to the present derivation. We draw attention to
the early single-hole results which show clear evidence for strong renormalization,
and effective sublattice-preserving hole hopping. Our conclusions are summarized in
section X. 
\\

\noin{\bf {II. Model and Symmetries}}
\\

The physics of no double occupancy is taken into account by 
representing the electron as a composite object created by  
$\cd _{i\sigma} = \bd _{i\sigma}h_i$, where $\bd _{i\sigma}$ creates a spinon of
spin $\sigma$ and $h_i$ destroys a holon at lattice site $i$, subject to the 
constraints that number of holons plus spionon at each site is one. 
Spin operators are represented as $S^+_i = \bd _{i\uparrow}b_{i\downarrow}$, 
and $S^z_i = \half (\bd _{i\uparrow}b_{i\uparrow} - 
\bd _{i\downarrow}b_{i\downarrow})$. Then the $t$-$J$ Hamiltonian is given by 
\beq H = - t\sum _{ij,\sigma} \cd _{i\sigma}c_{j\sigma} ~-~ 
 2J \sum _{ij}{\rm A}^{\dag}_{ij}{\rm A}_{ij},  \eeq
where the sum is over nearest neighbors,
${\rm A}_{ij} = \half \lbrack b_{i\uparrow}b_{j\downarrow} - 
b_{i\downarrow}b_{j\uparrow}\rbrack$ destroys a singlet connecting spinons
on nearest neighbor sites. The second term describes exchange interaction written
in terms of $\rm A_{ij}$'s, which makes the role of the singlets explicit. 
For the most part we consider the square lattice. However, later we will
need to include next-nearest-neighbor hopping within the plane and 
as well as hopping between planes.  

The Hamiltonian is invariant under a local $U(1)$ gauge transformation
\cite{bas1}: 
\beq b_{i\sigma} \rightarrow b_{i\sigma}e^{i\theta_i};~~~~
h_{i\sigma} \rightarrow h_{i\sigma}e^{i\theta_i}. \eeq
The number of spinons plus holons  
$$ {\cal N}_i = \bd _{i\uparrow}b_{i\uparrow} + 
\bd _{i\downarrow}b_{i\downarrow} + h^{\dag}_ih_i. $$
is then conserved at each site. The physical model lives in the projected
subspace defined by ${\cal N}_i = 1$, for each $i$. 
We can choose any statistics for spinons and holons
as long as the statistics of the gauge invariant quantities 
are correctly given. However, in an effective theory, in which spinons 
(and holons) are the low-energy excitations, we expect the system to choose a 
particular statistics.
\\

\noin{Half-Filled Case:}
\\

At half-filling the hopping term can be dropped. In the mean-field
approximation constraints are treated on the average, and singlets condense
so that the valence-bond order parameter $A_{ij} = <\rm A_{ij}>$ becomes nonzero. 
For dimensionality $d \geq 2$, the MF theory works well at low $T$ 
if spinons are bosons. For the symmetry of the order parameter we can choose: 
\beq A_{ij} = Ae^{i\half\vQ.(\vcr_i-\vcr_j)},\eeq 
where $\vQ$ is the zone corner wave vector. In $d = 2$, $\vQ = (\pi,\pi).$  
However, for $d \geq 2$ it has been shown that the ground state is not 
a pure RVB state since a fraction of spinons remains unpaired \cite{sar1,sar2}. 
Since spinons are bosons they condense independently so that
$<b_{i\sigma}> \neq  0$. This leads to a nonzero value for  
$<S^+_i> = <\bd _{i\uparrow}b_{i\downarrow}>$, 
giving rise to a two-sublattice antiferromagnetic order (in the x-direction). 

The ground state is therefore a mixture two competing phases,  
characterized by distinct order parameters, in which magnetization 
is reduced from its classical value by the presence of the singlet condensate. 
The MF theory describes the ground state properties in $d = 2$ rather accurately, 
and in $d = 3$, reproduces the results of Holstein-Primakoff quantum spin-wave theory 
\cite{sar1}. This success is remarkable, but not accidental, since   
the MF equations are identical to the equations describing the spin-wave 
theory \cite{aro}. In fact, the latter can be derived from the Schwinger-boson theory
by integrating out the constraints,
and expanding the action in powers of $1/S$. At finite $T$, AF order disappears
above $T_{AF}$ (in $d = 2$, $T_{AF} = 0$), but the RVB state survives up to 
$T_{RVB} > T_{AF}$. 
\\

\noin {Symmetry of the RVB State:}  
\\

In the MF approximation, condensation of singlets (and spinons) 
breaks gauge symmetry. Similarly, mobile holons lead to $<\hd_ih_j> \neq 0$
for $i \neq j$, which also breaks gauge symmetry. However, such local symmetries can not be 
spontaneously broken \cite{eli} since there is no rigidity with respect to 
phases of the order parameter $A_{ij}$ generated by the transformation (2). 
When averaged over these gauge equivalent choices, $A_{ij}$ 
(and other gauge-variant quantities) vanish. In this sense there is no condensation.
Similarly, spinon and holon Green's function are not gauge invariant, so these 
particles are not directly observable. The remaining phase fluctuations of the order 
parameter are studied via a gauge theory, as has been done by many authors.
However, there is still a transition 
across $T_{RVB}$ since the free energy is gauge invariant, and 
retains the characteristics of the condensate. In the uncondensed
state (i.e., above $T_{RVB}$) both phase and amplitude fluctuations are strong, 
and spinons and holons remain localized. In the condensed phase, fluctuations are weak. 
In this paper we will use the term condensation in this sense, and the term 
symmetry of a MF state to reflect the symmetry of the whole gauge-equivalent class. 

Furthermore, condensation can lead to certain type of gauge-invariant order on the lattice. 
For example, the RVB state has an important two-sublattice property (see below). Other
choices of symmetry lead to flux phases. Similarly condensation of spinon pairs and 
holon pairs lead to condensation of gauge-invariant composite entities such as a electron pair, 
whose symmetry depends on the symmetry of the constiuents. 

Although gauge fluctuations are supposed to be weak in the ordered phase at low $T$,
for a detailed quantitative theory (particularly for
transport properties) one has to include them \cite{iof}. Similarly, the
MF theory will not work well near $T_{RVB}$ where it will be necessary to
consider both amplitude and phase fluctuations. For the 2d
system these may be strong enough to convert the apparent second-order 
MF transition to a Kosterlitz-Thouless 
type transition (see \cite{ichi}). We will not study these issues here. 

At half filling, the MF theory has been formally extended by Read and Sachdev 
to situations where AF order is destroyed at $T = 0$ by quantum fluctuations \cite{read}.
In the ordered phase spinons are gapless, leading to gapless spin-wave
excitations. However, once magnetic order is destroyed, a gap
$\Delta _s$ appears in the spinon spectrum. Both cases can be described by the same MF theory 
with different values of $\lambda$, which acts as the spinon chemical potential.
The spinon spectrum has the form 
\beq \omega (\vk) = \lbrack \lambda ^2 - \phi (\vk)^2\rbrack^{\half},\eeq 
where
\beq \phi (\vk) = 4JA (\sin k_x + \sin k_y) \eeq 
is the spinon {\lq gap'} function. The minima of $\omega (\vk)$ are at 
$\pm (\pi/2,\pi/2)$. 
In the ordered state, $\lambda$ is chosen so that $\omega (\vk)$ vanishes at these
points leading two-sublattice AF order. 

In the absence of AF order, spinons are gapped since $\omega _{min} > 0$. 
Then, as shown by Read and Sachdev gauge fluctuations lead to a proliferation
of instantons, which results in the confinement of spinons into pairs \cite{read}.
In this case the Arovas-Auerbach type state is unstable relative to a 
valence-bond solid. In our case, we do not need to
worry about this, since there is no gap due to the presence of AF order. 
Away from half filling a spinon gap does appear,
but, as we shall see, holon motion causes singlets themselves to hop
around, which would destabilize the VB solid. 
\\

\noin A. Correlation Functions:
\\

Like the N\'eel state, the bosonic RVB state has a two-sublattice character of its own. 
From Eq.(5) we see that although $\phi (\vk)$
is not gauge invariant, it satisfies the gauge-invariant condition: 
\beq \phi (\vk) = \phi(\vQ -\vk). \eeq  
Now consider the spinon correlation function $B_{jm} = \half \sum _{\sigma} <\bd _{m\sigma}b_{j\sigma}>$ for any two sites $j$ and $m$. In the MF approximation this 
is given by
\beq B_{jm} = \frac{1}{N} \sum _{\vk} \cos (\vk .(\vcr _m - \vcr _j))\frac{\lambda}{\omega (\vk)}\lbrack 1/2 + n(\omega (\vk))\rbrack, \eeq
where $n(\omega)$ is the Bose function. Using Eq.(5), we find
\beq B_{jm} = B_{jm} \cos (\vQ.(\vcr _m -\vcr _j)). \eeq
Therefore, $B_{jm}$ vanishes if $j,m$ are on opposite sublattices. For ${j,m}$ on the same sublattice $B_{jm}$ is nonzero if the singlet condensate exists ($A \ne 0$). We will call this behavior {\em even}.
Similarly, consider the anomalous correlation function, $A_{jm} = \half <(b_{j\uparrow}b_{m\downarrow} - b_{j\downarrow}b_{m\uparrow}) >$, again defined
for any two sites $j,m$. (For nearest neighbors this is the RVB order parameter). It is
given by
\beq A_{jm} = \frac{i}{N} \sum _{\vk} \sin (\vk .(\vcr _m - \vcr _j))\frac{\phi (\vk)}{\omega (\vk)}\lbrack 1/2 + n(\omega (\vk))\rbrack. \eeq
From which we find
\beq A_{jm}= - A_{jm}\cos(\vQ.(\vcr _m - \vcr _j)). \eeq 
Thus $A_{jm}$ is zero if $j,m$ are on the {\em same} sublattice. For $j,m$ on opposite sublattices, $A_{jm}$ is nonzero as long as the singlet condensate exists. We will call this behavior {\em odd}. 
These symmetry properties are intrinsic to the RVB state since they hold even after long-range AF order
is destroyed, and for any $T < T_{RVB}$. They are also gauge invariant, even though the correlation
functions themselves are not, and therefore have observable consequences. Thus the spin-spin correlation function is given by ${\cal S}_{sp,ij} = <S^+_iS^-_j> =  - |A_{ij}|^2 + |B_{ij}|^2,$ which, as expected, alternates in sign. 

These properties imply that on average RVB singlets connect spinons on opposite sublattices. 
Therefore, the bosonic RVB state is similar to the short-range RVB state considered by Kivelson, Rokhsar and Sethna \cite{kiv}. What we have shown here is that this RVB state has the same red-blue property even in the presence of long-range magnetic order, i.e, even when magnetic correlations are not short-ranged.
\\

\noin B. Magnetically Disordered Phase
\\

Away from half-filling quantum fluctuations due to hole hopping is expected to destroy
long-range AF order at $T = 0$, leaving behind the RVB phase. Then bosonic
spinons acquire a gap $\Delta _s$, which is related to the magnetic correlation length \cite{aro,read}.
Near its minima the spinon energy 
has the form: $\omega (\vk) \approx [\Delta _s^2 + c_sk^2]$,
where $c_s= 2JA \sim T_{RVB}$, and $k$ is measured relative to the minima.
Then the correlation functions $B_{ij}$ and $A_{ij}$  behave as 
$$ B_{ij} \sim \frac{1}{r_{ij}}e^{-r_{ij}/2\xi}\cos (\half \vQ .(\vcr _j - \vcr _i)),$$
$$ A_{ij} \sim \frac{1}{r_{ij}}e^{-r_{ij}/2\xi}\sin (\half \vQ .(\vcr _j - \vcr _i)),$$
where $r_{ij} = |\vcr_i-\vcr_j|$, and $\xi = 2c_s/\Delta _s$ is the spin-spin correlation function.   

The spinon gap, and the consequent exponential decay of AF correlations
is the distinctive feature of the bosonic RVB state. In contrast, spinons are gapless
in a femionic (i.e., slave boson) RVB state, which would lead to power-law correlations.
Similarly, a Fermi liquid also has gapless spin excitations. Also, 
in the latter two cases, the magnetic correlations are peaked at a concentration-dependent
wave vector $\vq$, which equals $\vQ$ only at half filling. 
\\

\noin{\bf {III. Away from Half Filling: Early Work}}
\\

\noin {A. One-hole Physics and the $t^{\prime}-J$ Model
\\

Several authors have studied the behavior of a single hole moving in a
magnetic background by expanding in powers of hopping, and summing a selected a class of 
diagrams \cite{scm,kane}. It has been found that because of spin mismatch the hole can not hop 
coherently on the opposite sublattice, but it can
move coherently within the same sublattice. For large $t/J$, the 
bandwith corresponding to the coherent peak was found to scale with $J$, reflecting
strong renormalization of holes by spin fluctuations which makes the renormalized
hole large and heavy. The short-range incoherent hops due to $t$ leads to
an incoherent spectral background of order $t$. These results are supported by 
exact numerical work on finite lattices containing one hole \cite{dag}. They
can be used to estimate renormalized parameters for a many-hole theory.
However it is erroneous to construct a low-energy theory based on the one-hole spectrum, 
since a single hole does not destroy magnetic order the presence of which leads 
to a pole in the electron's Green function.

Motivated by the single-hole results, several authors have developed 
gauge theories for the many-hole system \cite{weig,lee3} based on the Schwinger boson 
RVB states (thus hoping to establish a connection with the Mott phase), with the 
assumption that the short-range incoherent hops caused by inter-sublattice
hopping ($t$) destroys AF order and strongly renormalizes the theory, 
leading to a sublattice-preserving $t^{\prime}- J$ model, 
where $t^{\prime} \sim J$ is a renormalized next-nearest-neighbor hopping parameter 
for the renormalized holes \cite{lee3}. While the assumptions are correct, the actual
renormalized Hamlitonian, as we show later, is quite different. 
\\

\noin{B. Spiral and Other Nearest-Neighbor States}
\\

One reason why Schwinger boson methods have not been taken seriously
is the fear of a spiral instability. For $x > 0$, moving holes are expected to 
rapidly destroy long-range AF order. As a physical hole hops, it carries 
its spin, creating a string of wrong (ferromagnetic) bonds, which cost energy and 
can not be easily repaired. 
In cuprates the AF insulating phase disappears beyond a small $x_c$,
giving way to a metal with no long-range magnetic order. 
Now, we can obtain a metallic state by extending the mean-field approximation
to the hopping term \cite{sar2}, which can be written as
$$ - t\sum _{ij}[ {\rm B_{ij}}^{\dag}{\rm D_{ij}} + h.c. ],          $$
where $\rm D_{ij} = \hd_jh_i$ and 
$\rm B_{ij} = \half \sum _{\sigma}\bd_{j\sigma}b_{i\sigma}$. In the MF 
approximation, in addition to singlets and spinons, the composite bosons 
created by $\rm B_{ij}$ and $\rm D_{ij}$ also condense, so that the hopping
term becomes
$$ \sum _{ij} B^*_{ij}\hd_jh_i + \sum _{ij,\sigma}D_{ij}\bd_{i\sigma}b_{j\sigma},$$ 
where $B^*_{ij} = <\rm B_{ij}>$ and $D_{ij} = <\rm D_{ij}>$, are the average 
spinon and holon backflows, respectively. Appropriate choices of symmetry for these
order parameters allow holons and spinons 
to hop coherently on to nearest-neighbor ($nn$) sites, which
mixes up the two sublattices. Different choices lead to different 
metallic states. The state with $B_{ij} = B$, $D_{ij} = D$, has spiral magnetic order with an
incomensurate wavevector. The spiral state is energetically favored over 
other relevant $nn$ states, e.g., the flux state, the ferrimagnetic states,
etc \cite{sar4}. Actually there is no sign of a spiral (or any other $nn$ state) 
has been found cuprates, or for that matter, in numerical treattments of the model. 

Incidentally, the slave-boson RVB theories are also based on 
$nn$ states, with a similar set of order parameters, although interpretations are 
different \cite{kot,ichi}. In both cases, the $nn$ states are very complex,
characterized by several order parameters ($A,B,D$ and $<b>$). 
This leads to a complicated phase diagram \cite{kot,ichi},
containing several metallic phases, more complicated than that seen in 
cuprates. Furthermore, since slave (or, Schwinger) bosons condense, the physical
electron Green's function has a pole, with a residue $\propto |<b>|^2$. Hence
the normal state is in effect a Fermi liquid at low $T$, with gapless spinon and 
holon excitations. 

Also, since slave-boson states do not work at half filling 
we should see a transition accompanied by a statistical transmutation upon
doping. To our knowledge, no 
experimental signature for such a transition has been found. 
Moreover, Haldane and Levine \cite{hal}, and independently, 
Read and Chakraborty \cite{read1} 
used Berry phase arguments to show that, in the presence of short-range RVB
pairing, spinons are bosons and holons are spinless fermions, as in our case. 
Finally, energetics also favor the Schwinger-boson representation,
for the simple reason that, given the same spectrum, fermions will have much
higher energy than bosons because of the Pauli exclusion principle. Therefore, 
bosonic spinons are better for the exchange term. Interestingly,
they are also found to be better for the hopping term for $x < 1/3$ 
even if $J = 0$ \cite{sar6}. 
\\

\noin{Instability of spiral states}

Moving holes induce ferromagnetic correlations, which compete with 
antiferromagnetism. In the spiral state the system compromises by tilting 
neighboring spins, so that the correlation is partially ferromagnetic.
While this allows the holons to delocalize and 
reduce kinetic energy, there is a substantial cost in exchange energy.
Therefore such a state may not be stable. Within a Hartree-Fock 
approximation, several authors have found an instabilty in the $2-d$ Hubbard model 
toward localization of holes in domain walls \cite{sch} for small $x$.
An instability toward phase separation was earlier found by 
Visscher \cite{vis}. 

By comparing the Hartree-Fock energies 
Hu {\etal} showed that spiral states are indeed unstable 
relative to both phase separation and insulating domain wall states in the 
entire region of the parameter space of interest \cite{hu}. 
The actual region of instability is likely to be even larger since contributions 
from RVB singlets to the exchange energy are not included in the Hartree-Fock treatment. 
Similar results were found for the $t-J$ model, where both antiferromagnetism 
and singlets are included \cite{sar4}. Since the spiral state has the lowest 
energy among the $nn$ states, it follows that none of these states are stable.

Longer range Coulomb interaction would destabilize the domain walls and spin-charge
separated states, and at very low doping could stabilze a Wigner crystal of holes.
However, since $t > J$ the localized states cost too much kinetic energy,
a metal must emerge at higher doping with or without such long-range 
repulsion. It follows that such a metal would have a lower energy than
spiral state, and can only appear via higher order hopping processes 
by renormalized holes, as indicated by the one-hole calculations.
To derive a renormalized Hamiltonian
and for the results obtained in this paper it is not necessary to
know the details of the renormalization procedure. But for the sake of consistency,
and for a more detailed theory, it is useful to consider the possible physical 
processes involved.
\\
 
{\noin {C. Destruction of Magnetic Order and Renormalization:}
\\

Our work is based on an earlier paper \cite{sar3} in which a 
self-consistent perturbation expansion in powers of hopping to one-loop 
order was used to study the destruction of magnetic order upon doping, 
which is expected to occur, but hard to show theoretically.
The point is that localization of electrons 
into spins (moments) costs considerable amount of kinetic energy. 
As discussed above, trying to 
recover the energy via spin-charge separated $nn$ states does not work. 
Another process which always exist is the binding of spin and charge into
physical electrons (or, holes). These try to propagate coherently and 
restore the Fermi liquid, which has low kinetic energy \cite{sar6}.  

In ref. \cite{sar3}, this process has been treated by RPA, in which the electron 
is treated as a collective excitation. The zeroth order Green's function 
$G_{c0}$ is simply the convolution of the spinon and holon Green's function. 
Summing the bubbles leads to the full electron Green's function \cite{sar3}
\beq G_c(\vk\omega) = \frac{G_{c0}(\vk\omega)}
{1 - \epsilon (\vk)G_{c0}(\vk,\omega)}, \eeq
where $\epsilon(\vk)$ is the energy of the noninteracting electron. 
This can be viewed as a generalization of the one-hole method 
where a similar expansion is carried out for one hole. 
In the many hole case, the effect of the
electron on the spinon and holon self-energies is also calculated,
using $G_c$. Unlike moving holons which only change the direction
of the local moments, a moving electron has a much more violent 
effect, as it also tends to delocalize, and hence destroy the moments.  
It was found that $x > x_c$, where $x_c$ is rather
small, long-range magnetic order is destroyed at $T = 0$, 
i.e., there is no Bose condensation of spinons \cite{sar3}.

It was also found that $G_c$ does
not have a pole at low frequencies, that is, the system is
not a Fermi liquid for small $x$. The reason is that in order to move 
an electron must avoid other electrons of opposite spin \cite{sar6}. 
For small $x$ and at low dimensionality, such pathways are rare. 
Consequently, the electrons (or, physical holes) do not become coherent. 
However, spinons become gapped since long-range magnetic order is destroyed. 
The gap shows up in $G_{c0}$ and hence in $G_c$ \cite{sar8}. In contrast,
in the one-hole case, the Green's function has a 
pole, which is not surprising since a single hole can not
destroy long-range AF order. In either case, the holes are
strongly renormalized by the short-range incoherent hops.
 
Interestingly, it was also found that the presence of a
RVB condensate leads to a second order hopping process in which 
a pair of holons hop together, accompanied by a singlet backflow, 
which automatically leads to superconductivity. This is the pairing
mechanism considered below. The theory was 
further developed in ref.\cite{sar5} where an effective 
Hamiltonian for the renormalized holons including the pairing 
term was derived. However, it was not entirely correct 
-- since the normal state was assumed to be a spiral metal,  
with nonzero $B$ and $D$, but without
long-range magnetic order. Such a state is not stable.  
Below we give a derivation which corrects this flaw.
\\

\noin{\bf{IV. Renormalized Hamiltonian for Small Doping}}
\\

The spinon gap allows us to obtain an effective Hamiltonian
involving renormalized holes and singlets, by integrating out the spinons
and setting $\omega = 0$ as was done in refs. \cite{sar3,sar5}. 
Here we give a simpler derivation using the usual continuity 
arguments \cite{lee3, kiv}. Since the $nn$ MF state 
is not stable we have $B = 0$ and $D = 0$. A moving hole affects 
antiferromagnetic configurations and singlets differently. In the
former case, a string of ferromagnetic bonds are created whose energy
increases with the length of the path, and which can not be easily
repaired quantum mechanically. The hole is essentially forced to return,
creating a renormalized hole localized within a bag of spin excitations. 
This process have been considered by many authors in the context of the 
one-hole problem \cite{scm,kane,dag}. It is also the reason for localization of 
holes in domain walls in the Hartree-Fock treatment of the Hubbard model \cite{sch,hu}. 
The hole moves rapidly within the bag,
gaining kinetic energy of order $t$. This would contribute an incoherent
(essentally, $\vk$-independent) background to the 
electron spectral function as well as to the optical conductivity. We assume that
disordering of the spins prevents long-range AF order, and leads to short-ranged
AF correlations. The renormalized hole is carries the spin excitataions, 
and is thus larger and more massive.  

After this intital renormalization, by continuity the
Hamiltonian would be similar to the original one, 
except for the absence of magnetic order,
as long as no additional symmetries are broken. It will be  
characterized by a hopping amplitude $t_{eff} < t$, and 
an exchange coupling $J_{eff}$. For small $x$, we expect $J_{eff} \sim J$. Also, 
one-hole calculations \cite{kane,dag}, suggest that $t_{eff}$ scales with $J$ for $t > J$.
Such estimates can be used since the renormalization process
involoves only nearby hops, and therefore at low-hole densities it is
essentially a one-hole problem. An important aspect of the one-hole calculations, 
is that the local constraints are imposed exactly. The constraints
obeyed by the renormalized particles are thus expected to be weaker, although
the renormalized Hamiltonian retains gauge invariance.

The singlets are affected less drastically by hole motion since the
offending configuarations can be repaired more easily, as discussed below. 
We therefore assume that the problem is already renormalized by the fast 
processes described above. 
When a renormalized hole hops from sublattice $a$ to $b$ it breaks a singlet 
and creates two spinons, costing, say, an energy $\Omega > 2\Delta _s$. 
The system relaxes by a second hop, after which the singlet is reconstructed.  
In the low frequency limit ($\omega << \Omega$) these processes can be described by  
\beq H_{int} = - \frac{t_{eff}^2}{\Omega}<{\cal P} \sum _{jl\sigma,mi\sigma^{\prime}}
\cd _{m\sigma}c_{i\sigma}\cd_{l\sigma ^{\prime}}c_{j\sigma ^{\prime}} {\cal P}>, \eeq
where $c_{i\sigma} = \bd_{i\sigma}h_i $; but now all the operators correspond
to renormalized particles. Here $<{\cal P}...{\cal P}>$ means
that we keep only those terms that involve renormalized singlets and holes. Although
nominally similar to superexchange, the intermediate state does not involve double
occupancy which has already been projected out. As shown below the new interaction terms can
be rewritten in terms of singlet operators ${\rm A_{ij}}$. Now, $A_{ij}$ and $B_{ij}$ decay 
exponentially as $e^{ - r_{ij}/2\xi}$ \cite{read}, where $\xi$ is the magnetic correlation length. 
Therefore, to obtain a minimal Hamiltonian, it is sufficient to retain only the short-range 
hopping terms since longer range terms are exponentially suppressed. The longer-range
hopping terms also preserve the underlying symmetries, and neglecting them
should not change the physics qualitatively.

Once the singlet is broken, there are three ways to remove the excess energy. 
First, the hole can hop back, and the singlet reconstructed. 
This is confining and its effect is to renormalize the chemical potentials. The remaining
two processes can lead to hole propagation.

\noindent A. {\em One-hole process}: The hole hops to another site on 
sublattice $a$, and the singlet is reconstructed on a different link [Fig. 1]. Notice that
this process involves successive hopping by two electrons of opposite spin, but only one holon. 
Projecting onto the spinon singlet subspace [${\cal P} (\bd_{i\sigma}b_{j,-\sigma}{\cal P}
= \sigma {\rm A}_{ij}$, etc], we obtain
\beq -  \frac{t_s}{2}\sum _{ijl}{\rm A}^{\dag}_{jl}{\rm A}_{ij}h^{\dag}_ih_l(1 - h^{\dag}_jh_j), \eeq
where, ${\rm A}^{\dag}$ creates a singlet and $\hd$ creates a {\em renormalized} holon,
and $t_s = 4\frac{t_{eff}^2}{\Omega}$ is the effective intra-sublattice hopping parameter.
Now, the term $(1- h^{\dag}_jh_j)$ arises from the intermediate excited state. It is
gauge invariant, and for small $x$, can be approximated by its average value $1-x$, leading to
\beq -  \frac{t_s}{2}(1-x)\sum _{ijl} {\rm A}^{\dag}_{jl}{\rm A}_{ij}h^{\dag}_ih_l, \eeq
This term describes hopping by a holon within the same sublattice, accompanied by a 
backflow of spinon singlets. 
\begin{figure}[htbp]
\centering
\includegraphics[height=2cm,width=8.0cm,angle=0]{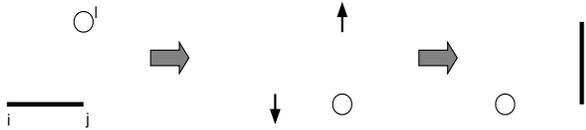}
\caption{One hole process. A hole hops from $l$ to $j$, breaking the singlet $(ij)$ denoted by the solid line. Here $i$ and $l$ are on sublattice $a$ and $j$ is on sublattice $b$. The hole then hops to $i$, and the singlet is reconstituted at $(jl)$}
\end{figure}   

The one-hole term above differs from the usual next-nearest-neighbor 
hopping ($t^{\prime}$). In the early RVB-gauge theories based on similar RG
considerations it was {\em assumed} that the result of second order hopping is to
generate an effective $t^{\prime}$ term of the form 
$$ t^{\prime}\sum _{il}\cd_{i\sigma}c_{l\sigma} + h.c., $$
where $i$ and $l$ are next nearest neighbors. To generate such a term from Eq. (12)
would require consecutive hops by two electrons of the {\em same} spin. This is not easily
done when the initial and final states belong to the singlet subspace, as in our case.
For after the first hop the remaining electron will have opposite spin. Then to generate
a $t^{\prime}$ one has to flip both spins via additional higher order intermediate
processes, which are not very likely. (It can be done via $J^{\prime}$ interaction.
But for this model $J^{\prime}$ is usually much smaller than $J$). Of course the actual
model that applies to real cuprates will have a $t^{\prime}$ term to start with which,
as will see later, will play a role since it also preserves the
two-sublattice character, but it would not replace the dominant 
one-hole term derived above.

B. {\em Two-hole process}: The system also relaxes if a second hole
hops from sublattice $b$ to $a$, and the singlet is reconstituted on a different link [Fig 2]. 
This process yields a term
\beq  - t_s\sum _{ij;lm} {\rm A}^{\dag}_{ml}{\rm A}_{ij}\hd_i\hd_jh_lh_m, \eeq   
which describes hopping by a singlet, accompanied by the backflow of a holon pair. 
This is the small $x$ form of the interaction derived earlier \cite{sar5}, but here the
normal state is different because of the one-hole term above.
\begin{figure}[htbp]
\centering
\includegraphics[height=2cm,width=8.0cm,angle=0]{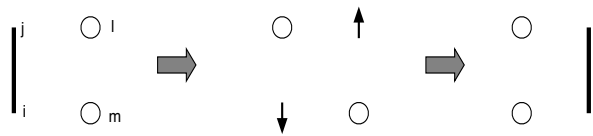}
\caption{Two hole process. A hole hops from $l$ to $j$, breaking the singlet $(ij)$, 
a second hole hops from $m$ to $i$, and the singlet is reconstructed at $(ml)$.}
\end{figure}  

The full Hamiltonian also includes the renormalized exchange term, characterized by
$J_{eff}$. For the purposes of this paper 
it is sufficient to treat $J_{eff}$, $t_{eff}$ and $\Omega$ 
as phenomenological parameters. However, as discussed earlier, 
one-hole calculations \cite{dag} suggest that they all scale with $J$.
Hence the renormalized Hamiltonian describes a strong-coupling problem. 
For cuprates, $J \sim 0.1 eV$, and is the largest energy scale in the 
renormalized problem. If numerically $t_{eff} > \Omega$ then 
a frequency cut-off can be used. By continuity
qualitative physics will not be changed. 
As it turns out (see below), actually the effective holon 
hopping amplitude $t_{eff}A$ is $\sim JA  \sim T^*$. Since 
$A$ is only a small fraction of unity and decreases with $x$, a cut off
may not be necessary, at least for some values of $x$.

Because of the spin gap our renormalized Hamiltonian describes a 
short-range RVB model for small $x$ \cite{kiv}. However, the SR Hamiltonians used
by earlier authors \cite{rokh} were different (for example, the second term which 
is responsible superconductivity was absent) since they were not required 
to agree with Arovas-Auerbach state at half filling. Also, holons were thought to be
bosons. It was later shown that for short-range RVB models
they are actually fermions, which agrees with our choice \cite{hal,read1}. 
Our renormalized Hamiltonian can not possess a conventional Fermi liquid state 
since the latter has gapless spin (and charge) excitations. 
\\

\noin{\bf {V. Analysis of the Renormalized Hamiltonian}}
\\

\noin{A. Non-ordered Phase: Strange Metal or Insulator?:} 

There is always a state in which no symmetry is broken (even at the MF level).
Then singlets are not condensed,i.e., $<A_{ij}> = 0$. Hence, 
$$<\hd _ih_j> = 0$$
for $i \ne j$, so that there is no coherent propagation of holons.
Holons are localized due to strong gauge fluctuations. 
This state has no long-range order, and would
always occur at sufficiently high $T$ (above $T^*$ in our case). 
We identify it with the so-called strange metal phase of the cuprates. Since
electrons are also incoherent, there are no coherent charge carriers at all in 
this state. Therefore, the strange metal is not a metal; it is
a new type of quantum insulator. 
In effect, this state corresponds to a Gutzwiller
projected Fermi sea; but one with strong singlet fluctuations, which lead to the
gapping of the spinons (and, hence, the electrons), and non-Fermi liquid behavior. 
In contrast, in other RVB theories the state above $T^*$ is metallic, 
with gapless spin and charge excitation \cite{kot,ichi}. In these theories the
nonordered state appears at a much higher temperature, and is ignored. 
\\

\noin{B. Pseudogap Phase}
\\

To proceed further, we first do a MF decomposition of $H_{int}$ to
obtain two separate Hamiltonians for spinons and holons, which breaks
gauge symmetry. Then, from the one-hole term, we get  
\beq -  \frac{t_s}{2}(1-x)\sum _{ijl} \lbrack <{\rm A}^{\dag}_{jl}{\rm A}_{ij}>h^{\dag}_ih_l
 + <h^{\dag}_ih_l>{\rm A}^{\dag}_{jl}{\rm A}_{ij}\rbrack. \eeq
In principle, we can 
find a state for which $<A_{ij}> = 0$ (no singlet condensation), but
$<{\rm A}^{\dag}_{jl}{\rm A}_{ij}> \neq 0$. Such a state will not be favored
energetically, since the energy gained $O(t_sx)$ per bond is low compared with 
$J_{eff}A^2$ gained from singlet condensation. 

This leaves the RVB state which becomes
stable below $T^*$ with $<A_{ij}>  \neq 0$. From Eq. (2) we obtain 
$$<{\rm A}^{\dag}_{jl}{\rm A}_{ij}> = A^*_{jl}A_{ij}
 = - A^2e^{i\half\vQ.(\vcr_i + \vcr_l)}.$$ 
Using this in (15) we obtain an effective holon hopping Hamiltonian 
\beq H_{h0} = - \sum _{ij} t_{h,ij}h^{\dag}_ih_j, \eeq
which describes coherent holon propagation within the same sublattice.  
Here $t_{h,ij} = \pm t_h$ for $i,j$ next-nearest neighbors along the $(1,\pm 1)$ diagonal
direction, $t_{h,ij} = t_h/2$  for next-next-nearest neighbors (along the x and y directions),
and $t_h = t_sA^2(1-x)$.  The holon energy is then given by 
\beq \epsilon_{h1}(\vk) =  - 2t_h + 2t_h(\sin k_x + \sin k_y)^2. \eeq
The holon band (hence, metallic conduction) appears as soon as $A \ne 0$ without
additional symmetry breaking. As expected, the spectrum has the two-sublattice
character since $\epsilon _{h1}(\vk) = \epsilon _{h1}(\vQ - \vk)$. 

We can do a similar MF decomposition of the two-hole term (Eq. 14), and
replace the $\rm A_{ij}$'s be
their expectation values, which yields a holon-holon interaction term of the form
\beq H_{h,int} = - t_sA^2 \sum _{ij;lm}\lbrack {\rm F}^{\dag}_{ij}{\rm F}_{ml} + h.c.
\rbrack, \eeq 
where ${\rm F}^{\dag}_{ij}= \hd_i\hd_j$ creates a holon pair on the link ${ij}$, and the 
sum is over plaquettes.
The order of the indices follows from the symmetry of $A_{ij}$ and is very important.

Once the holons are propagating, the pair-hopping term will also 
contribute to the normal state holon spectrum. 
Doing a MF decomposition of $H_{h,int}$ we obtain a term
$$H_2 = t_sA^2\sum _{ij,ml}\lbrack D_{li}\hd _{j}h_m + D_{mj} \hd _ih_l   - D_{li}D_{mj}  
+ h.c. \rbrack, $$  
where  
\beq D_{li} = <\hd_ih_l> = \frac{1}{N} \sum _{\vk} e^{i\vk.(\vcr_l - \vcr_i)}f(\epsilon _h(\vk) - \mu), \eeq 
is the average holon hopping amplitude, and $f$ as the Fermi-Dirac function. 
Using the symmetry of the holon spectrum, we find that
$D_{li} = 0$, when $l$ and $i$ are nearest-neighbors. And, $D_{li} = \pm D_1$
for $l$ and $i$ along the $(1,\pm 1)$ diagonal directions, respectively, with
\beq D_1 = - \frac{1}{N}\sum _{\vk} \sin k_x\sin k_y f(\epsilon _h(\vk) - \mu). \eeq
Therefore the extra diagonal hopping terms have the same symmetry as the original ones.
Furthermore, since the minima of the holon spectrum is at $\pm \half (\pi,-\pi)$,
$D_{1}$ is positive. Using the fact that hopping along $(1,1)$ is accompanied by a 
backflow along $(1,-1)$ the extra diagonal terms are found have the same signs 
as before. Including these the full holon spectrum becomes
\beq \epsilon _h =  - 2t_h + 2t_h(\sin k_x + \sin k_y)^2 + 4D_1t_s \sin k_x\sin k_y. \eeq
The parameter $D_1$ has to be calculated self-consistently.
However, for small $x$, the main contribution comes from near the minima of the spectrum (
$\vk = \pm (\pi/2,-\pi/2)$), where we can set $\sin k_x\sin k_y = - 1$, which yields $D_1 \approx x$. 

So far we have considered only the holon part. The motion of holons (or holon pairs) 
is accompanied by a backflow of singlets. The mean-field singlet Hamiltonian 
is then given by $H_b = H_J + H_{b1} + H_{b2}$, where $H_J$ is the usual exchange
term (see Eq. 1), with an exchange constant $J_{eff}$;
\beq H_{b1} =  -  \frac{t_s}{2}(1-x)\sum _{ijl} {\rm A}^{\dag}_{jl}{\rm A}_{ij}<h^{\dag}_ih_l>, \eeq 
is the singlet backflow term associated with the one-hole process, and 
\beq H_{b2} = - t_s\sum _{ij;lm} {\rm A}^{\dag}_{ml}{\rm A}_{ij}<\hd_i\hd_jh_lh_m>, \eeq   
is that associated with the two-hole process. The last two terms only appear 
below $T^*$. They will contribute to the singlet condensation energy in the pseudogap phase.
They are suggestive of the emergence of coherent singlet excitations, which are spin-$0$, 
chargeless vector bosons. These will contribute to low-T specific heat, and thermal conductivity, 
but not to electronic transport. We will study them in a future paper. 
\\

\noin{Symmetry Properties of the Pseudogap Phase}
\\

The pseudogap state preserves the two-sublattice (red-blue) character of the RVB state.
To see this consider the correlation functions for the renormalized particles. 
Using $\epsilon_h(\vk) = \epsilon_h(\vQ - \vk)$, we find $D_{ij} =  <\hd_jh_i> = D_{ij} \cos (\vQ.(\vcr _j - \vcr _i))$, which is thus nonzero only on the same sublattice (even). The spinon correlation functions, and
hence the gauge-invariant magnetic correlation functions have the same symmetry as in the Mott phase. 
The electron hopping amplitude $P_{ij,\sigma} = <\cd _{i\sigma}c_{j\sigma}>= -B_{ij}D_{ij}$. Since $B_{ij}$ and $D_{ij}$ are both even, $P_{ij}$ is nonzero only on the same sublattice. The Fourier transform of $P_{ij}$
is the momentum distribution function, which thus satisfies $n_c(\vk) = n_c(\vQ -\vk)$. In a Fermi liquid
the electron hopping amplitude shows a power law decay. In our case, the holon hopping amplitude $D_{ij}$ 
shows a metallic, i.e, power-law behavior since holons are gapless. However, since spinons are gapped $B_{ij}$, 
the full electron hopping amplitude $P_{ij}$, decays exponentially, reflecting non-Fermi liquid behavior. 

The metallic character of the holons can also be seen in the gauge-invariant
charge structure factor. Let $\rho _i = \hd_ih_i - <\hd_ih_i>$ measure the excess hole density. Then the charge structure factor is given by: ${\cal S}_{ch,ij} = <\rho _i\rho _j> =  - |D_{ij}|^2,$ for $i \ne j$, and ${\cal S}_{ch,ii} = x(1 - x)$.  This is nonzero on
the same sublattice, and exhibits the long-range oscillatory structure of a metal. In the momentum space it
satisfies ${\cal S}_{ch}(\vk)  = {\cal S}_{ch}(\vQ - \vk)$ (Fig. 3). In contrast, ${\cal S}_{ch}(\vk)$ of an ordinary metal increases from zero at $\vk = 0$ and becomes a constant for $q > 2k_F$. In our case, an image of the behavior near $\vk = 0$ appears near $\vk = \vQ$.
\begin{figure}[htbp]
\centering
\includegraphics[height=6cm,width=6.0cm,angle=0]{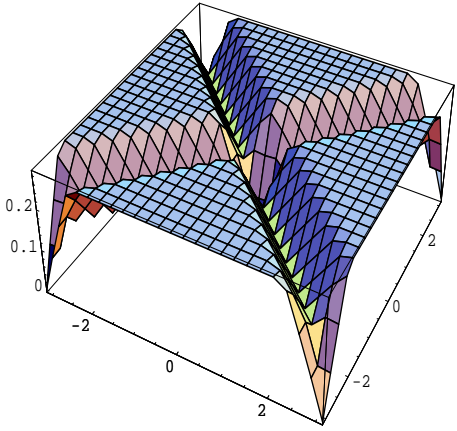}
\caption{Charge structure factor, showing thetwo-sublattice symmetry. The calculation is approximate. 
Note the symmetry between  $\vk = 0$ and $\vk = (\pi,\pi)$, which is distributed among the four corners of the Brillouin zone. The valley region is elevated relative to $\vk = 0$. In a normal metal, the structure near $(\pi,\pi)$ is absent.}
\end{figure}  
Experiments probe the bare correlation functions. These are dominated by short-range incoherent processes that do not preserve the two-sublattice property, which is therfore not easy to see. The best candidate is ${\cal S}_{ch}(\vk)$ since holon motion is coherent. The experimental ${\cal S}_{ch}(\vk)$ would no longer vanish at $\vQ$, but there will still be a dip. 
\\

\noin{\bf {VI. Superconducting State}}
\\

It is clear from the structure of the two-hole term (Eq. 19)
that the system can lower its kinetic energy if two holons hop as a pair. 
Pairs will condense, 
leading to $F_{ij} = <{\rm F}_{ij}> \ne 0$. Let us define the pairing order parameter for
physical electrons as a singlet
\beq {\cal C}_{ij} = 
<(c_{j\downarrow}c_{i\uparrow} - c_{j\uparrow}c_{i\downarrow})/2>. \eeq 
Then, ${\cal C}_{ij} = - A_{ij}F^*_{ij} \ne 0$ since the spinons are already condensed,
giving rise to superconductivity below $T_c \le T^*$. The order parameters $A_{ij}$
and $F_{ij}$ are not gauge invariant, but ${\cal C}_{ij}$ is. The symmetries of
$A_{ij}$ and the holon spectrum are of course already known. 

We solve the superconductivity problem by a mean-field approximation. 
Since holons are fermions the holon pairing order parameter satisfies $F_{jm}= - F_{mj}$.
We denote the vector $F_{jm}$ by two components: $F_{jm} = iF_x$ 
if $m$ is to the right of $j$, and $F_{jm} = iF_y$ if $m$ is above $j$ in the 
$y$ direction. The prefactor $i$ is chosen so that order parameter for the electron (Cooper)
pair is real. For a uniform system, we can take $|F_{jm}| = F_0$, 
but the phases along $x$ and $y$ need not be the same. 
Without loss of generality we can choose, $F_x = F_0$ 
and $F_y = \alpha F_0$, with $\alpha = e^{i\theta}$. 
The choice of $\alpha = \pm 1$ leads to ${\cal C}_{x} = \pm {\cal C}_{y}$, 
corresponding to s-wave (d-wave) symmetry for the electron pair wave function. But other
values of $\alpha$ can, in principle, lead to $s + id$ or $s + d$ symmetries.
To find the correct symmetry we compare the free energies for different choices.
The MF Hamiltonian is given by
\beq H_{MF} = \sum _{\vk} \xi (\vk) \hd _{\vk}h_{\vk} + 
\half \Delta _h(\vk)(\hd _{\vk}\hd _{-\vk} + h.c. ), \eeq   
where $\xi (\vk) = \epsilon _h(\vk) - \mu$, $\mu$ is the holon chemical potential,
and $\Delta _h(\vk) = 2t_s F(\vk)$ is the holon gap function, with
\beq F(\vk) = 2F_0 (\sin k_x + \alpha \sin k_y). \eeq  
The holon spectrum $\epsilon _h(\vk)$ is assumed to contain the Hartree contribution from the
interaction term. The Hamiltonian is diagonalized by the Bogulyubov transformation.
The self-consistent equations for the order parameters are then given by
\beq F_{\eta} = \frac{4t_s}{N}\sum _{\vk} \sin k_{\eta}(\sin k_x + \alpha \sin k_y)
\frac{\tanh (\beta E(\vk)/2)}{E(\vk)}, \eeq
where $\eta = (x,y)$, and 
$$E(\vk) = \lbrack \xi ^2(\vk) +  |\Delta _h(\vk)|^2 \rbrack ^{1/2}.$$
Since $F_x$ is real we find that solutions exist only for $\sin \theta = 0$, i.e,
for $\alpha = \pm 1$. Combining the equations for $F_x$ and $F_y$, we obtain
\beq \frac{1}{t_s} = \frac{1}{N}\sum _{\vk} W(\vk) (\sin k_x + \alpha \sin k_y)^2
\frac{\tanh (\beta E(\vk)/2)}{E(\vk)}, \eeq
$W(\vk)$ is a suitably chosen cut-off function.
\\
\begin{figure}[htbp]
\centering
\includegraphics[height=6cm,width=6.0cm,angle=0]{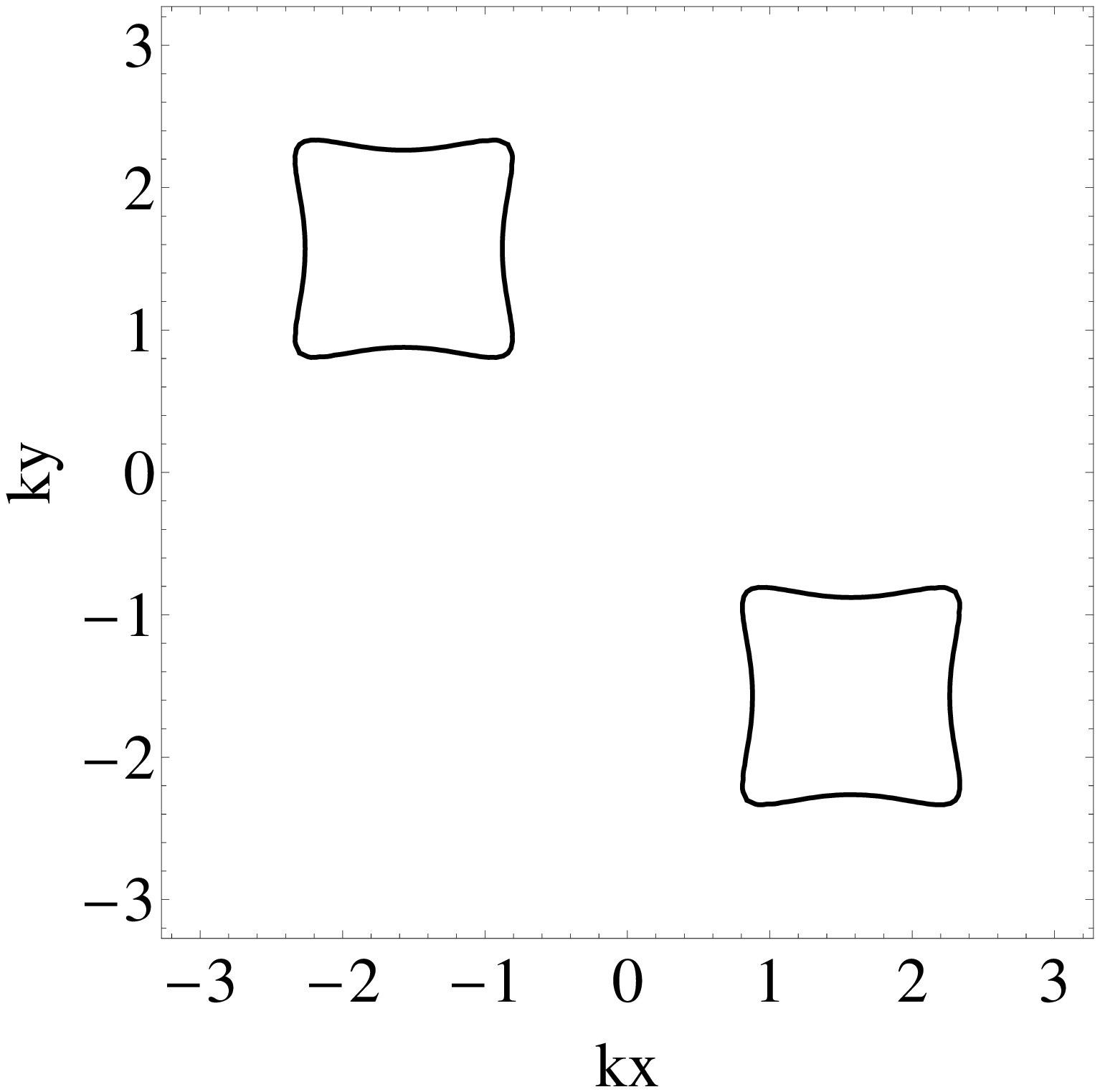}
\includegraphics[height=6cm,width=6.0cm,angle=0]{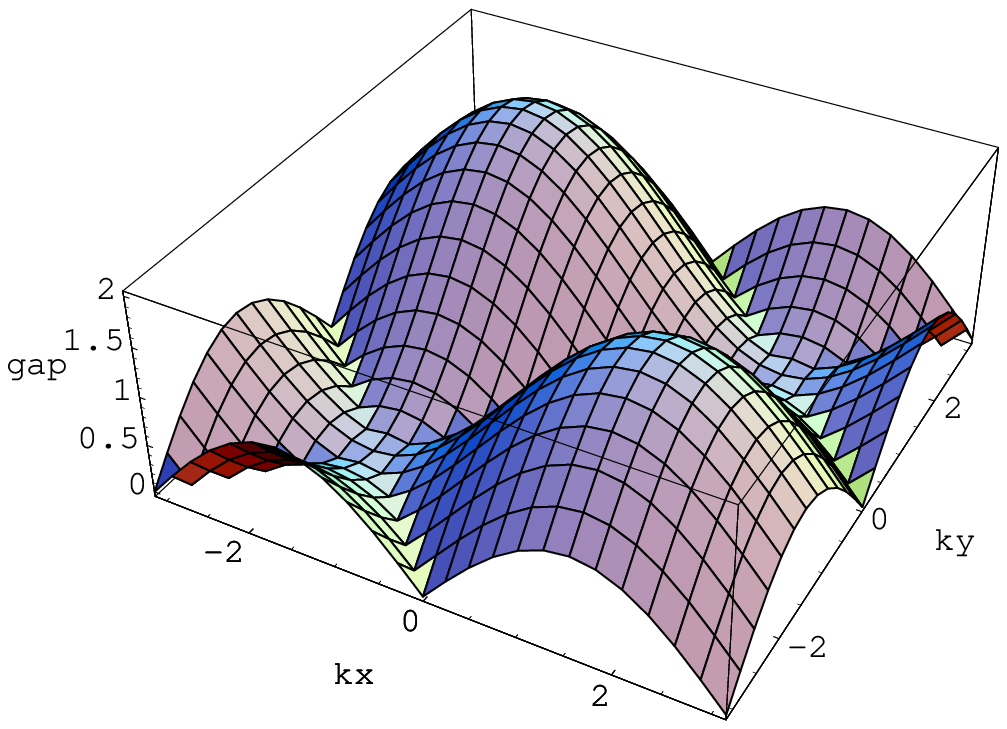}
\caption{Origin of $d$-wave symmetry. Upper panel: Holon Fermi surface. Holons live within the pockets centered at $\half(\pi,-\pi)$ and $\half(-\pi,\pi)$. The shape of the FS is somewhat different when compared with ref.\cite{sar9} because the holon spectrum has corrections from the two-hole term. The FS is not directly
observable since under a gauge transformation it is moved and deformed. 
Lower panel: The symmetry factor $|\sin k_x + \alpha \sin k_y|$. It has broad maxima in the hole-rich region for $\alpha = - 1$, resulting in maximum condensation energy. In contrast, the symmetry factor vanishes in this region for  $\alpha = 1$ ($s$-wave). Under a gauge transformation both the Fermi surface and the symmetry factor move together to preserve these gauge-invariant results.}
\end{figure}  
To find the symmetry let us consider the $T = 0$ case, for which $\tanh \beta E(\vk)/2) = 1$. We
have solved this equation numerically with and without a cut-off function. Quite generally, we find that
$\alpha = -1$, (i.e., $d$-wave) is favored, as it leads to the largest $F_0$, and hence the 
largest condensation energy. The origin of this result can be understood from the following simple
considerations. The dominant contribution to the sum comes from the region where $|\xi (\vk)| = |\epsilon _h(\vk) - \mu_h|$ is small, and the symmetry factor $|\sin k_x + \alpha \sin k_y|$ is large. As shown in Fig. 4, the holon Fermi surface is in the second and fourth quadrant, exactly where $\sin k_x + \alpha \sin k_y$ has maxima for $\alpha = - 1$ ($d$-wave) and vanishes for $\alpha = 1$ ($s$-wave)). Hence, $d$-wave always wins. Thus the symmetry of the superconducting order parameter is determined by the symmetry of the underlying RBV state.

Note that since $F(\vk) = F(\vQ - \vk)$, the two-sublattice property is also preserved in the superconducting state. The holon pairing function is odd since for any two $i,j$ satisfies $F_{ij} = - F_{ij}\cos \vQ.(\vcr_i - \vcr _j)$. Since
$A_{ij}$ is also odd, the electron pairing function ${\cal C}_{ij} = - A_{ij}F^*_{ij}$ is also odd. Similarly, the symmetries of $n_c(\vk)$ and spin-spin correlation function remain unchanged in the superconducting state. 
The charge structure factor, however, picks up an additional contribution: ${\cal S}_{ch,ij} = |F_{ij}|^2 - |D_{ij}|^2$, and is no longer restricted to the same sublattice; but like the spin-spin correlation function, it oscillates in sign.

Note, the renormalized Hamiltonian describes a strong-coupling problem since
the one-hole and two-hole terms are essentially governed by the same energy scale $t_s$. 
The holon wave functions depend on $x$; but not on $t_s$. Hence 
the BCS like MF approximation is expected to work only for $T << T_c$. In particular, treatment of the 
transition region, and the calculation of $T_c(x)$ will require strong-coupling techniques. 
Similarly, MF approximations would not work near $T^*$, where fluctuations will be strong. 
The phase diagram shown in Fig. (5)
is thus only schematic. The main point is, unlike earlier RVB theories, the number
and type phases for small $x$ are similar to those in cuprates. Moreover, as discussed
below, the qualitative behavior of various phases can be easily understood.  Given the
numerous unusual properties of the cuprates, and the 
highly constrained nature of the renormalized Hamiltonian, the predictions can be
put to severe experimental tests. So far we have found no contradictions.
\\

\begin{figure}[htbp]
\centering
\includegraphics[height=8cm,width=8.0cm,angle=0]{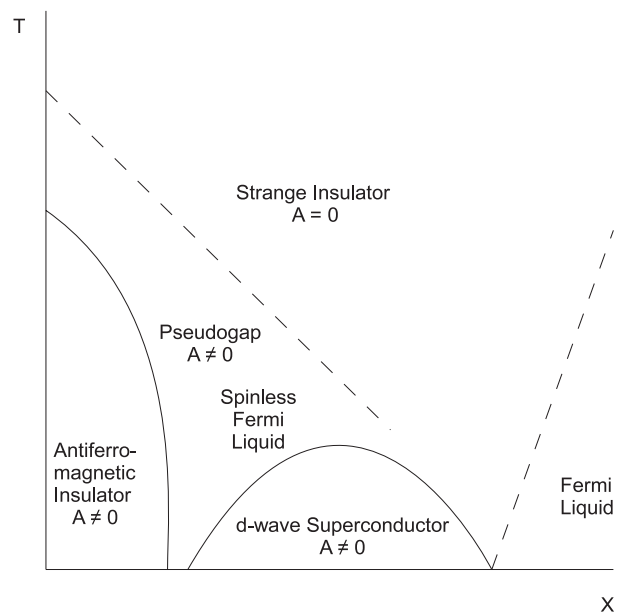}
\caption{Schematic phase diagram. The state above $T^*$ is a novel insulator and is unordered. 
The pseudogap metallic state is a spinless Fermi liquid. Execpt for the strange insulator,
the remaining phases are characterized by RVB order parameter $A \ne 0$}
\end{figure}  

\noin{\bf{VII. Comparison with Experiments:}}
\\

Clearly, the present theory is consistent with the important features of the experimental
phase disgram. By construction, it is consistent with the physics at half filling.
The normal state has spin-charge separation and is not a Fermi liquid. There is a transition 
from a strange phase to a pseudogap metal below $T^*$. As discussed later in section VIII, 
these phases are two-dimensional, whereas the superconducting state exhibits 3d behavior. 
However, the phases themselves are predicted to have novel chracterisitcs that defy
conventional wisdom. The key ones are: (1) there is a true spin gap $\Omega$ which exists in all three
phases, and which is distinct from the pseudogap that appears below $T^*$, (2) holons 
are confined above $T^*$, so that the strange phase is an insulator rather
than a metal;  (3) below $T^*$ holons form a spinless Fermi liquid of concentration $x$, 
without an observable (small) Fermi surface. The pseudogap metal and the supercondor are 
characterized by the RVB order parameter and the 
two-sublattice property, whereas the strange phase does not have any order.
These results are robust as they follow from symmetry and particle statistics.
A large number of important conclusions can be drawn
from the very structure of the effective Hamiltonian in each phase,
which do not require detailed calculations, and which can be tested against
experiments. We now examine recent experimetal data to
show that there is very strong evidence in
support of these predictions. We separate the experimental findings 
into three groups: the spin sector, the charge sector and the electron sector. 
\\

\noin{A. The Spin Sector:}

The spin gap implies the existence of an additional temperature scale $T^0 \sim \Omega/k$. 
In the pseudogap region the AF correlation length  $\xi _{spin}$ is expected to be short 
-- no more than a few lattice spacings -- due to strong renormalization processes 
implicit in our theory. Then, $\Omega \sim  J_{eff}(x)$. On the other hand, 
$T^* \sim J_{eff}A < T^0$, since $A$ is only a fraction of unity (from the MF theory). 
At half filling, RVB ordering arises from long-range singlet-singlet interactions 
mediated by gapless spinons. In the doped region this interaction is weaker and 
short-ranged because of the gap. However, now
the renormalized hopping terms favor singlet condensation since 
holons can propagate coherently only if singlets are correlated. 
The qualitative behavior is easy to understand. 

Above $T^0$ we expect free spins to dominate. Then the uniform paramagnetic 
susceptibility $\chi _{para}$ should be Curie like, i.e., decrease with increasing $T$. 
Below $T^0$ singlets will form, and $\chi _{para}$ will start decreasing with 
{\em decreasing} $T$. Note that there is no transition at $T^0$; it is just a broad 
crossover scale. Hence, we can associate it with
the maximum of $\chi_{para}$. With decreasing $T$, there will be more 
singlets and they will become increasingly more correlated, and eventually condense 
at $T^*$. Below $T^*$, $\chi _{para}$ will decrease much more rapidly, and vanish as 
$T \rightarrow 0$ (because of the gap), even in the absence of superconductivity.
The magnetic behavior above $T^*$ is determined by 
fluctuating singlets correlated over a distance $\xi _{singlets}(T)$, together 
with weakly correlated free spins. The behavior near $T^0$ and above, where singlets
are weakly correlated, can be crudely understood in terms of a one singlet (two-site) problem, 
for which $\chi _{para}$ is given by
$$\chi_{para2} = \frac{\mu ^2\beta}{3 + e^{\beta J_{eff}}}. $$ 
This has gap $\Omega = J_{eff}$, and it vanishes exponentially as 
$e^{-\beta J_{eff}}$ as $T \rightarrow 0$, shows a Curie like $1/T$ 
behavior at large $T$ with a maximum at $T^0 \approx 0.62 J_{eff}/k$.   
Fig. 6 shows $\chi _{para2}$ as a function of $T$.  
\\
\begin{figure}[htbp]
\centering
\includegraphics[height=6cm,width=8.0cm,angle=0]{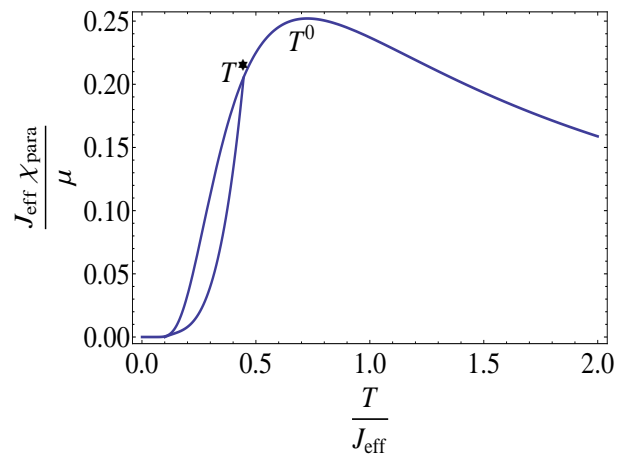}
\caption{Paramagnetic susceptibility for a two-site problem. Note the maximum
at $T^0$, the spin-gap scale. The line on the right below $T^*$ is
schematic, showing how singlet condensation at $T^*$ for a many-site problem would further
depress $\chi _{para}$.}
\end{figure}  

The pseudogap has been observed in cuprates by using nuclear magnetic resonanance (nmr) 
techniques \cite{war,wal}. There is a rapid decline in $\chi _{para}$, as measured by the Knight
shift, below $T^*$, which is far above $T_c$. A similar downturn is seen in the spin-lattice
relaxation rate, $1/T_1$. More interestingly, the higher crossover temperature $T^0$ 
was also found in many of these materials (for a review, see \cite{tim}). 
That $T^0$ has not been seen in some materials is 
more a matter of not going to high enough temperatures. For example, $T^0$ was not
seen in initial measurements on YBa$_2$Cu$_4$O$_8$ upto 400 K.  Curro {\etal}\cite{cur} 
extended the measurement to $715 K$ and found a broad maximum at $T^0 \sim 500 K$, with
$\chi _{para}$ decreasing slowly on both sides of $T^0$. In this material
$T^*$ is about $240 K$, and $T_c \sim 81 K$. It is interesting to note that 
$T_0$ and $T^*$ vary roughly the same way, decreasing with increasing $x$,
consistent with the theoretical expectation that they are proportional to
the same energy scale $J_{eff}(x)$. 

Our theory also predicts that $\chi _{para}$ would vanish in the normal state as 
$T \rightarrow 0$. This is harder to confirm since supercondcutivity intervenes. 
However, as discussed in ref. \cite{tim}, in highly underdoped systems such as
YBCO6.7 \cite{wal} and YBa$_2$Cu$_4$O$_8$ \cite{zim}, there is almost no sign of a 
superconducting transition in the nmr data at $T_c$, indicating that there is no 
additional spin pairing at $T_c$, although $\chi _{para}$ continues to decrease.
Similar results have also 
been seen in moderately underdoped Bi1222 \cite{wal} by Walstedt {\etal} who
argue that the lack of any effect at $T_c$ indicates that the gap 
is unrelated to superconductivity, and represents spin-charge separation. 
\\

\noin{B. Electron Sector: Tunneling DOS}

Further support for the theory comes from tunneling experiments, 
which probe the electron spectra (charge plus spin). 
Unlike magnetic properties, tunneling DOS \cite{ren,miya} 
shows a sizable effect near $T_c$ in underdoped Bi2212. For $T << T_c$, there 
is a gap in the DOS, with well-developed conductance peaks
at $\pm \Delta _{sc}$, except that $2\Delta _{sc}/kT_c$ is much larger than the
BCS value. As $T$ approaches $T_c$ from below, 
conductance peaks decrease in height and disappear at $T_c$, as does
the zero-bias peak originating from the Josephson current. This is what is
expected. However, while the gap decreases as $T_c$ is approached, it  
does not close; instead a gap like structure with a depressed DOS continues to exist above
$T_c$. Although, sometimes it is referred to as the pseudogap
in the tunneling literature, the gap actually exists far above $T^*$ (as
measured in the nmr experiments), that is, it also exists in the
strange phase up to high temperatures, i.e., it is the spin gap.
Such a behavior can not be understood in terms of $nn$ (e.g., slave-boson states, 
or a Fermi liquid. 

The tunneling data are clearly consistent with our theory in which
the spin gap exists in all three phases. The gap and the
singlet condensation qualitatively account for the nmr data in the normal state.  
Superconductivity arises from a pairing of spinless holons. Since there is
spin-charge separation $\chi _{para}$ and $1/T_1$ would largely be unaffected below 
$T_c$, as seen in the experiments. However,
tunneling experiments measure the DOS of the electron, containing
both spin and charge. At the MF level this is a
convolution of spinon and holon DOS. Above $T_c$ holons are gapless, 
so the gap in the electron Green's function reflects the spin gap. 
This is why tunneling gap exists far above $T^*$. Below $T_c$, 
a gap also opens up in the charge spectrum, as holons pair; 
this explains the increase in the {\em total} gap size with 
decreasing $T$, and the appearance of the conductance peaks.
Also, experimentally the $T = 0$ gap increases with decreasing $x$, 
mirroring the behavior of $T^0$ associated with the much larger spin gap,
as expected from the theory.

Experimental observation of a spin gap far above $T^*$ rules out a mechanism based 
on preexisting electron pairs, carrying both spin and charge. To account
for the gap, such pairs should be formed near $T^0$, and presumably 
condense below $T_c$. It is then hard to explain $T^*$. There are other
difficulties. The condensation
at $T_c$ must increase the binding energy of the Cooper pairs, enough 
to cause the large depletion observed in the DOS below $T_c$, and the
observed increase in the size of the gap. However, since electrons forming a
Cooper pair also carry spin, such a large increase will be accompanied by
a large change in $\chi _{para}$ and $1/T_1$ below $T_c$, a behavior not
seen in the experiments \cite{wal}. 
\\

\noin{C. The Charge Sector}
\\

Strange Insulator: A key prediction of our theory is that holons are confined (nonpropagating) 
above $T^*$ since $A = 0$. Furthermore, since electrons are incoherent and gapped 
the {\lq strange metal'} is not a metal, but a new type of insulator. 
A strong evidence for this comes from the measured in-plane dc resistivity, 
$\rho$. In a metal $\rho $ is supposed to saturate at the Mott value. 
Actually, while the linear $T$ dependence of the  $\rho (T)$ has received most of the 
attention, it has been pointed out by many authors that $\rho $ shows no sign of saturation,
and in the underdoped regime, far exceeds the Mott value \cite{taka,seg,tak}. 
The fact that $\rho (T)$ increases with $T$ does not make the system a metal since
$\rho$ also increases in a disorder-induced (Anderson) localized insulator as long as 
inelastic mean-free path $\ell _{inel}(T)$ is less than the localization length, 
$\xi_{loc}$. A quantitative theory is beyond the scope of this paper; but
the qualitative behavior can be understood, as follows. 
In the case of Anderson localization $\xi _{loc}$ is independent of $T$; then 
$\rho (T)$ would show an upturn at low $T$, as $\ell _{inel(T)}$ exceeds $\xi _{loc}$. 
In our case, the appropriate localization length is the distance over which holons can
move coherently, which is the singlet-singlet correlation length $\xi _{singlet}(T)$
(not the shorter AF correlation $\xi _{spin}$).  Since $\xi _{singlet}(T)$ increases
with decreasing $T$, and becomes infinite at $T^*$ (at the MF level) where $\ell _{inel}$
is finite. Hence there can not be a low-$T$ upturn, which is consistent with experiments. On the other
hand, since there is no coherent holon motion in the perpendicular direction
the c-axis resisitivity should show an upturn, as seen.

A clear indication is also obtained from the frequency-dependent conductivity
$\sigma (\omega)$ which, for a metal, should exhibit a Drude peak. But in our
case there would be no such peak above $T^*$. Recent
experiments in LSCO \cite {tak} and B21212 \cite{san} show that such a 
peak does not exist for small $x$. Instead, one has nearly a flat spectrum
over a range of frequencies of order 1 eV. The lack of a Drude peak 
indicates the absence of coherent charge carriers. The
flat spectrum would come from incoherent hopping. At higher
doping, a broad peak develops, which is not surprising since in this case the
physical electron comes into play as one approaches the Fermi liquid. 
\\

\noin {Pseudogap State: Emergence of Metallic Conduction}
\\

The theory predicts that below $T^*$ holons become coherent and form a spinless 
Fermi liquid of concentration $x$. Therefore $\rho(T)$ should drop 
rapidly with decreasing $T$, and become metallic. Indeed, experimentally $\rho (T)$ 
is found to drop faster than linearly \cite{taka,buch,bat} below a temperature 
which approximately agrees with $T^*$ obtained from nmr experiments \cite{tim,bat}. 
At lower T, $\rho(T)$
becomes metallic and has a residual impurity contribution \cite{ito,seg,tak},
as expected of a metal. At low-$T$, $\rho (T) \propto T^2$ 
due to fermion-fermion scattering \cite{taka,ando}, also as expected.  
For LSCO one also finds that $\rho$ does show an upturn at very low $T$, which
is believed to be due to disorder-induced localization \cite{taka}.

Recent experiments \cite{ando,pad} in underdoped LSCO and YBCO show that the Hall 
coefficient $R_H$ is independent of $T$ and proportional to $1/x$ in the pseudogap
regime, as expected of a metal. This should be compared with the strong T-dependence 
in the 'strange' insulator phase. With increasing temperature, $\rho(T)$ is seen to
deviate from the $T^2$ to eventually a linear-$T$ behavior above $T^*$.

Strong evidence also comes from optical conductivity $\sigma (\omega)$, which shows a 
Drude peak at low $T$, with an integrated area (spectral weight) $\propto x$ \cite{orn,pad}.
It is characterized by a small plasma frequency, consistent with a small bandwidth,
or heavy holons, as in our theory. With increasing $T$, $\sigma(\omega)$ broadens,
but at a rate too large to be attributed to thermal effects
in a Fermi liquid, and merges into the incoherent background above 
$T^*$ \cite{san,tak,pad}. 

These transport properties, taken together, strongly
support the view that the charge carriers are holes, rather than electrons,
and they form a Fermi liquid below $T^*$, and when combined with the nmr data,
they suggest that these holes are spinless, supporting our view. They 
also suggest that at small $x$ there are no coherent charge carriers above $T^*$.
\\

\noin{\em Lack of a Fermi Surface:} 

If the pseudogap state is a Fermi liquid of
concentration $x$ as the transport experiments suggest, where is the corresponding, 
presumably, small Fermi surface? This is a real puzzle. Attempts to understand this \cite{ando} 
in terms of the `Fermi arc' found in the photoemission experiments \cite{ino,yos} does not 
make sense for many reasons. First, there are no quasiparticles associated with the 
Fermi arc. The electron Green's function is completely incoherent in the normal state; 
there are no sharp peaks in the photoemission spectrum. Therefore, the Fermi liquid is not
formed by electrons. Furthermore, as discussed earlier, neither the mmr data
nor the gap in the tunneling spectra is consistent with an electron Fermi liquid.

In contrast, our theory predicts that the holon Fermi surface is unobservable, 
even in principle. This is because the holon Green's
function, and thus the holon spectrum, are not gauge invariant. Hence the holon Fermi 
surface can be deformed and moved around by a gauge transformation. 
The holon Fermi surface does not exist when averaged over all the gauge-equivalent 
copies. This is why experiments do not see it.

Another prediction is that since $\chi _{para}$ decreases rapidly below $T^*$, and eventually
vanishes as $T \rightarrow 0$, the {\em total} uniform magnetic susceptibility of pseudogap
metal would become more diamagnetic with decreasing $T$ due to the orbital motion of
holons. The effect should be small near $T^*$, but should become more 
prominent in the metallic region (at lower $T$).  
Unfortunately, superconductivity intervenes. Note that the diamagnetic response discussed
here should not be confused with strongly $T$-dependent response found by
Wang (\etal} \cite{wang} considerably above $T_c$. The latter is presumably due to pairing
above $T_c$. Such strong-coupling effects are not ruled out by our theory, but its
treatment would require  more sophisticated techniques. We point out that in the anlaysis of 
Wang {\etal} the 
normal-state diamagnetic contribution appear to have been subtracted out along
with other weakly temperature-dependent terms. Therefore more experimental work
will be needed to see the effect in the normal state.

In our theory superconductivity appears via a pairing of holons.
Since both terms in the effective Hamiltonian arises from hopping, 
condensation energy is $\propto - t_s^2/t_h$, which causes a reduction in the 
kinetic energy, as observed \cite{deut,mole}. Our mean-field treatment 
implies that spin-charge separation continues to exist in the superconductor.
For small x, there is some evidence for this since
$\chi _{para}$ is essentially unaffected at $T_c$; 
but more work is needed to clarify this issue.
\\
 
\noin{\bf{VIII. Origin of Two Dimensionality}}
\\

In this section we consider the origin of two-dimensionality of the
normal state(s), which is central to the occurrence of high-$T_c$ 
superconductivity in cuprates. These are highly anisotropic 3d materials.
Ordinarily the metallic
state would show 3d behavior below some $T_{\perp}$ since confinement within
a plane would cost too much kinetic energy. For cuprates, this  
would imply $T_{\perp} < T_c$, which is small at low doping,
and actually vanishes at some critical $x$. In other words, the normal state 
appears to remain two-dimensional as $T \rightarrow 0$ 
(in the absence of superconductivity). On the other hand, the other
phases in cuprates: the AF state at half filling, 
the superconducting state, and the Fermi liquid state at large $x$ are all
three dimensional, as one would expect.   

In theoretical studies, the 2-d nature of the normal state is 
usually {\em assumed}, not established. One simply analyzes the 2d 
Hamiltonian. The question arises: does the corresponding normal state 
remain two-dimensional when hopping is turned on in the perpendicular direction? 
If it does not, the theory should be discarded. This requirement, seldom tested, 
provides a powerful constraint on any theory.  

Out of plane hopping can be modeled by adding
$$ - t_{\perp} \sum _{i,z} \lbrack \cd_{i\sigma}(z)c_{i\sigma}(z+1) + h.c \rbrack, $$
to the original model, where $t_{\perp} << t$; and for simplicity we consider a 
tetragonal lattice. Suppose we put $U = 0$, and treat the anistropic hopping model 
as an effective model for the Fermi liquid state. 
Obviously, this would show 3d behavior. The same is true if the particles are bosons, 
as in a superconductor or a quantum magnet. Similary, the $nn$ (e.g., the spiral 
or the slave-boson) states should be three-dimensional since they are modeled by 
similar effective MF Hamiltonians, in which fermions and bosons hop 'freely' in all
three directions. The issue of confinement has been
studied for coupled 1d Luttinger chains which have gapless excitations \cite{cla}.

In our theory the gapping of the spin-excitations provides a protective mechanism
for 2d confinement. This is far from obvious since, as we have seen, holons can
delocalize in the plane even though spinons are gapped. So why not in the $z$ 
direction? First, note that a nonzero $t_{\perp}$ leads to an exchange interaction 
$J_{\perp} = 4t_{\perp}^2/U = J(t_{\perp}/t)^2 << J$. Since a spin can belong
to only one singlet at a given time it follows that the spin
singlets are formed only on the x-y plane in the ground state. In order to form them
along the $z$ direction one has to break singlets in the 
plane, which will cost an energy of order $J - J_{\perp}$ per singlet, {\em as long
as spin excitations are gapped}.

Now, suppose a hole (or a projected electron) hops on to the adjacent layer, 
breaking a singlet, which creates two unpaired spinons, one in each plane. 
The cost in energy is $\sim \Omega$, which can not be removed by the hole hopping on
to a third layer since that would create more unpaired spinons costing more
energy. Now, there are only two processes by which the system can relax. The
hole can hop back to the original layer, 
which means that the normal state is two dimensional. The second one is the two-hole 
process similar to the one 
discussed earlier, that is, a second hole follows the first. As in the plane,
the pair hopping is accompanied by a singlet backflow. Therefore 
superconductivity is three-dimensional, which explains the 3D enhancement of
$T_c$. 

Although this interlayer {\em holon pair} hopping is nominally simlilar to the interlayer pair
tunneling mechanism \cite{and3}, the physics is quite different. In
the ILT theory electron pairs tunnel. It plays the primary role;
there is no intra-later pair hopping. The spinon spectrum is gapless, the interlayer hopping 
matrix is diagonal in $k_{xy}$, i.e., long-ranged. In our case, the hopping
is localized in real space. The primary mechanism for superconductivity
is the intra-layer holon pair hopping, which contibutes the
the main part of the condensation energy. The interlayer hopping 
process makes superconductivity three-dimensional and enhances $T_c$.
\\

\noin{Superconductivity in the 3D system}
\\

The second process described above contributes the following interplane pair hopping term
\beq H_{inter} =  - t_{zs}\sum _{ij,z}\lbrack 
{\rm F}^{\dag}_{ij}(z){\rm F}_{ij}(z+1)+ h.c \rbrack, \eeq
where $t_{zs} > 0$ is an effective pair hopping amplitude which is ($<< t_s$), the
intraplane pair hopping amplitude. In this paper we will treat it as free parameter.
Note that the vector fields ${\rm F}_{ij} = <h_jh_i>$ lie on the plane. 
Furthermore, this term does not modify the normal holon spectrum at the Hartree-Fock level.

The MF problem for the superconducting state is again reduced to a 2D problem 
with an additional pairing term. 
The equation for the order parameter (with $d$-wave symmetry)
now becomes 
\beq \frac{1}{t_s + t_{zs}} = \frac{1}{N}\sum _{\vk} (\sin k_x - \sin k_y)^2
\frac{\tanh (\beta E(\vk)/2)}{E(\vk)},  \eeq
where, as before,
$$E(\vk) = \lbrack \xi ^2(\vk) +  |\Delta _h(\vk)|^2 \rbrack ^{1/2},$$
but with a modified gap function
\beq \Delta _h(\vk) = 4(t_s+ t_{zs})F_0 (\sin k_x - \sin k_y). \eeq  
The equation for $x$ is unaffected. As
shown in Fig. (7) $F_0$ increases with $t_{zs}$, for given values of $t_h$,
the effect increasing slowly with increasing $x$. Note that the sign of $t_{zs}$ is
determined by the symmetry of the underlying RVB state, and is
of considerable importance since, as seen from the figure, $F_0$
actually decreases if the sign is reversed by hand (bottom line).
\\

\begin{figure}[htbp]
\centering
\includegraphics[height=8cm,width=8.0cm,angle=0]{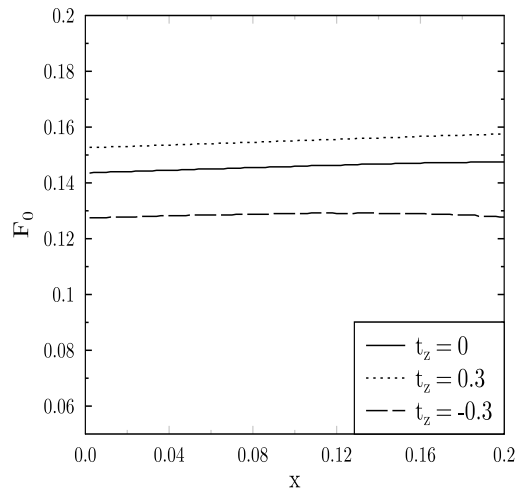}
\caption{3D enhancement of superconducting condensation energy: 
Order parameter vs interplane
pair hopping amplitude $t_z$, which is the same as $t_{zs}$ in the text.
It is enhanced in the actual case (dotted line). However, it 
is reduced if the sign of the interaction is reversed by hand (bottom line).}
\end{figure}  

\noin{\bf{IX. Electron doping vs hole doping}}
\\

High-$T_c$ superconductivity also occurs in electron-doped systems. This is expected
theoretically since the square-lattice Hubbard (or $t$-$J$) model has an electron-hole symmetry 
about half filling. The symmetry is realized by the transformation: 
$c_{i\sigma} \rightarrow \cd _{i\sigma}$ on one sublattice and 
$c_{i\sigma} \rightarrow - \cd _{i\sigma}$ on the other. If we assume that the orbital
structure in the CuO$_2$ plane is roughly the same for the two cases, an electron-doped system at 
an electron concentration $1 + x$ should exhibit roughly the same behavior as a hole-doped system at
a concentration $1 - x$. Experimentally this is not found to be the case. In the electron-doped
system the superconducting state occupies a much smaller region in the 
parameter space, and $T_c$'s are also much smaller. The AF insulator phase
occupies a much larger region.

In a real system there will always be an asymmetry since one
has to include an intra-sublattice hopping term in the original model:
\beq H^{\prime} =  t^{\prime}\sum _{il}\cd_{i\sigma}c_{l\sigma} + h.c., \eeq
where $i$ and $l$ are next-nearest neighbors, and this term 
changes sign under the e-h transformation. A rough estimate 
from ARPES and electronic structure calculations is: 
$t^{\prime}/t \approx (0.1 - 0.3)$ \cite{pav}. The $t^{\prime}$ term would 
generate a $J^{\prime} = 4{t^{\prime}}^2/U$, which is very small 
since $J^{\prime}/J = (t^{\prime}/t)^2$, and can be neglected.  
In general, $t^{\prime}$ should not change the physics qualitatively; and
for a Fermi-liquid or a $nn$ state the quantitative difference should not be too
large since physics is dominated by the $t$ term.	The observed difference is 
much larger than what is expected.

In our case the difference can be significant since $t$ is renormalized away,
and the hopping parameter $t_h$ that characterizes renormalized
intra-sublattice hopping (see Eqs. 17-18) is only $\sim J$. It should
be noted this term {\em does not} change under the e-h transformation. Therefore,
the effect of the $t^{\prime}$ is to increase the net intra-sublattice hopping
amplitude in one case and decrease it in the other. 
Of course, $t^{\prime}$ will also be renormalized to $t^{\prime}_{eff}$, but
RG effect will be smaller since this term moves an electron within the same 
sublattice which has ferromagnetic correlations.
Now, in the MF approximation $H^{\prime}$ becomes
\beq H^{\prime} = - t^{\prime}_{eff}\sum _{ij} \lbrack 2B_{ij}\hd _jh_i 
+ D_{ij}\sum _{\sigma}\bd _{i\sigma}b_{i\sigma} + h.c. \rbrack, \eeq
which generates intra-sublattice holon hopping. Here
$B_{ij} = <\bd _{i\sigma}b_{j\sigma}>$ and $D_{ij} = <\hd_jh_i>$ are the average hopping 
amplitudes for spinons and holons along the diagonal. Above $T^*$ this term does not
contribute since $B_{ij} = 0 = D_{ij}$. 
We assume that the $t^{\prime}_{eff}$ is small enough that it does not by itself
break gauge symmetry and generate coherent hopping of holons above $T^*$.

The situation is quite different below $T^*$, since the diagonal holon hopping amplitue
$D_{ij} \neq 0$ in the pseudogap metal; and the diagonal $B_{ij}$ is also nonzero in the presence
of the singlet condensate. The symmetry of $B^*_{ij}$
is easily calculated from the MF theory, and we find $ B^*_{ij} = \mp B_2$, along the
diagonal $(1,\pm 1)$ directions, where
\beq B_2 = \frac{1}{N} \sum _{\vk} \sin k_x \sin k_y \frac{\lambda}{\omega (\vk)}
\lbrack 1/2 + n(\omega (\vk))\rbrack. \eeq   
Since the spinon spectrum has minima at
$\vk = (\pi/2,\pi/2)$, $B_2 > 0$. Thus the $t^{\prime}$ term contributes 
\beq \epsilon^{\prime} _h(\vk) = - 4t^{\prime}_h \sin k_x\sin k_y, \eeq
to the holon spectrum, where $t^{\prime}_h = t^{\prime}_{eff}B_2$.
Therefore, the extra holon hopping term in the diagonal direction
has the same symmetry as the earlier one; except that the sign is opposite for 
the hole-doped case for which, $t_h^{\prime} > 0$.
As shown in Fig. 8 (upper panel), the extra term has maxima at $\pm(\pi/2,-\pi/2)$,
where the original spectrum is a minimum, making the band shallower and holons heavier
(and increases the holon DOS). 
Since $t^{\prime}$ term does not contribute to the pairing
interaction, the net effect is to reduce pair breaking, which increases $F_0$ and $T_c$. 
This is seen in Fig. 9 where we plot $F_0$ vs $x$ at $T = 0$.

\begin{figure}[htbp]
\centering
\includegraphics[height=6cm,width=6.0cm,angle=0]{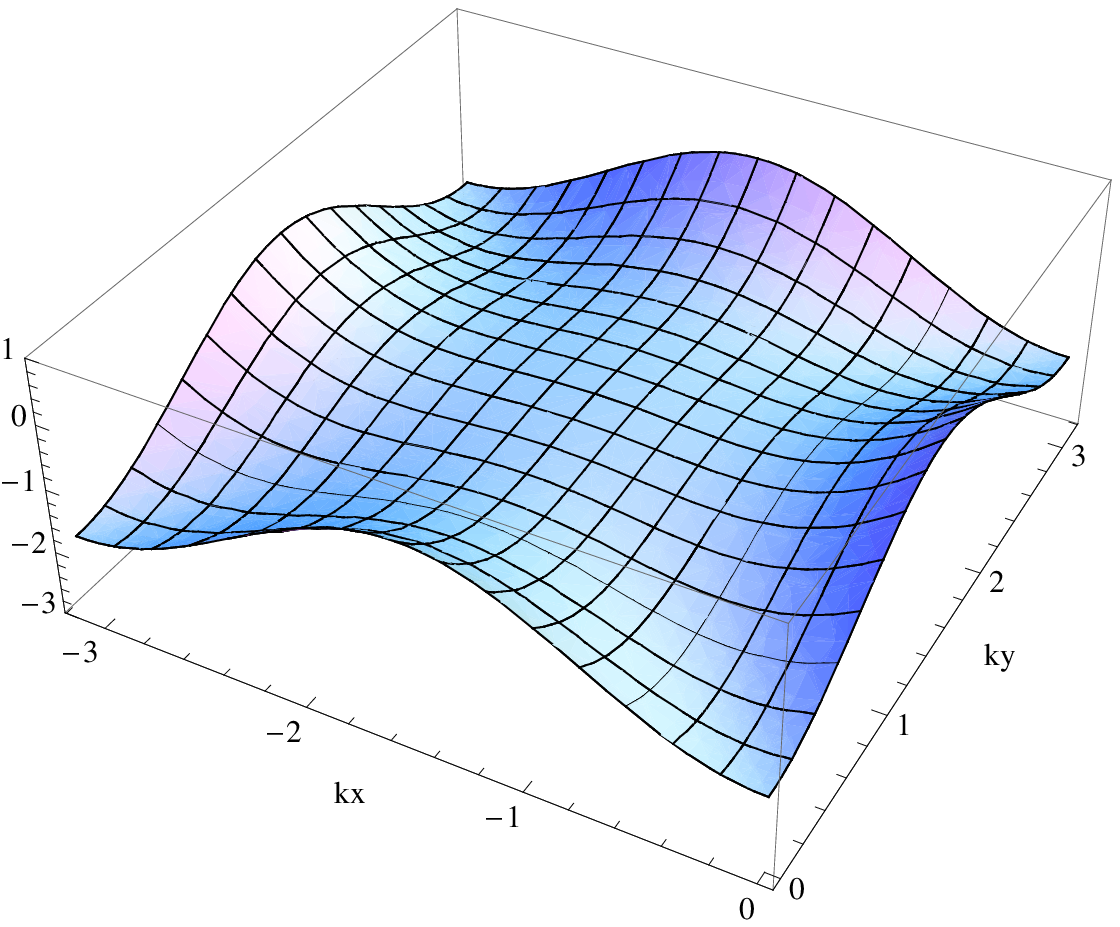}
\includegraphics[height=6cm,width=6.0cm,angle=0]{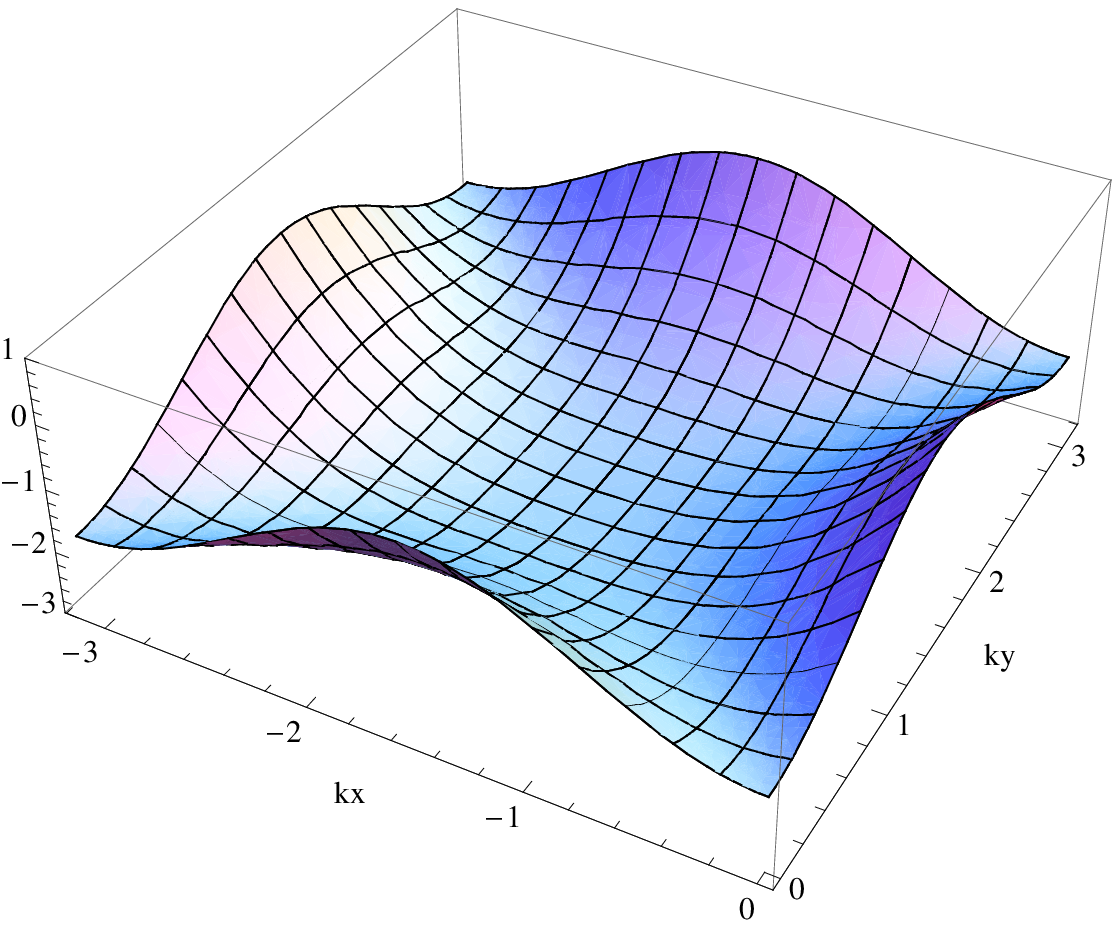}
\caption{The origin of electron-hole asymmetry: the holon spectrum in the second quadrant 
in the presence of next-nearest-neightbor hopping $t_{eff}^{\prime}= \pm 0.2t_{eff}$. For 
positive $t_{eff}^{\prime}$ (hole-doped case) the minimum becomes shallower (upper panel), 
which would to lower Fermi velocity and higher $F_0$. 
For the electron-doped case, $t_{eff}^{\prime}$ 
is negative, and minimum deepens (lower panel), which would lead to higher fermi velocity, 
and lower $F_0$.}  
\end{figure}  

The problem of an electron-doped system at an electron density of $1 + x$  
is the same as that of a hole-doped system at a hole density of $x$, except that
the sign of $t^{\prime}$, (and hence, that of $t_h^{\prime}$) is now changed. In this
case, the total holon hopping amplitude along the diagonal add. The extra term
has minima at the minima of the original spectrum (Fig. 8 - lower panel), 
making the band steeper, which making holons lighter and thus
decreases the DOS. As shown in Fig. 9, $F_0$ is now smaller, as one would expect. 
We stress that these effects are amplified
because the pairing interaction and holon hopping are of the same
order of magnitude.

\begin{figure}[htbp]
\centering
\includegraphics[height=8cm,width=8.0cm,angle=0]{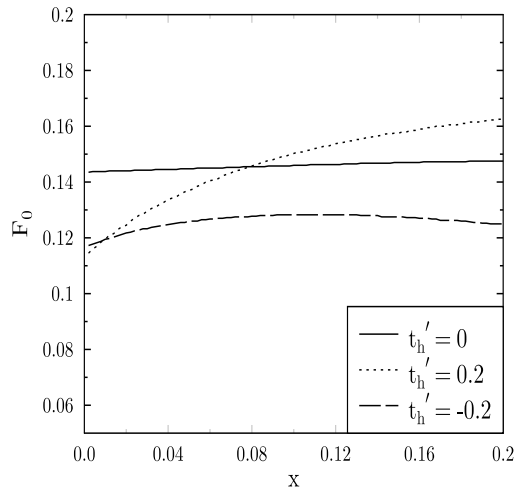}
\caption{Electron-hole asymmetry: the condensation energy vs the effective 
diagonal hopping
amplitude $t^{\prime}_h$ for fixed $x$. The hole doped case (dotted line) corresponds to  
$t^{\prime}_h > 0$, and the electron doped case (broken line) corresponds to $t^{\prime}_h < 0$}
\end{figure}  

Interestingly the theory also provides a straightforward explanation for why
the size of the AF region is larger for the electron-doped case simply on the
basis of the symmetry of the RVB state. The point is that average spinon hopping
amplitude $B_{ij}$ measures {\em ferromagnetic} correlations between sites $i$
and $j$. It is nonzero when $i$ and $j$ are on the {\em same} sublattice, and has the 
largest value
when the system has long-range AF order. This can be verified
from Eq. (35) which shows that right-hand side is largest when the 
spinon energies $\omega (\vk)$ are smallest, i.e., when the spectrum is gapless
(corresponding to long-range AF order). 
We also see from Eqs. (20,21) that $D_{ij} = \pm D_1$, with $D_1 > 0$ along
the $(1,\pm 1)$ directions. Hence, the contribution of the $t^{\prime}$ term to the
energy per bond is $ t_{eff}^{\prime} B_2D_1 $
which has the same sign as $t^{\prime}$. It follows that AF ordering will be opposed by
this term for the hole-doped case ($t^{\prime} > 0$), and will be favored for the 
electron-doped case ($t^{\prime} < 0)$. 
\\

\noin{\bf {X. Conclusion}:}
\\

In this paper we have shown that assumptions of continuity from half filling and 
renormalization gives
rise to the number and type of phases that are seen in the experimental phase diagram.
The theory makes three major predictions: a spin gap which plays a pivotal role in all 
three phases; a spinless Fermi liquid below $T^*$, but one without
without a Fermi surface; and a state above $T^*$ which is not a metal but a novel 
quantum insulator. Except for superconductivity, the motion of holons do not introduce 
any new order, other than those at half filling.
The normal states are predicted to be rather different when compared with 
other theories. We have presented a careful review of the old and new 
experimental results in the charge, spin and electron sectors, and found very strong
evidence in support of these predictions. 

Superconductivity also occurs naturally, via pair hopping 
of holons, driven
by the existence of pre-paired spinons. It is found that the possible symmetries of the 
superconducting order parameter are severely restricted because the symmetry of the spin 
part of the electron pair is already determined at half filling. We have shown that this
leads to a robust $d$-wave. Remarkably, the theory provides a natural explanation
for the two-dimensionality of the normal state, and also of the pronounced difference
between hole-doped and electron-doped cuprates. Most of results follow from the type and 
symmetry of the renormalized Hamiltonians, detailed calculations are not necessary.

The theory provides a basis for future calculations of finite-temperature properties, 
which would require careful treatment of fluctuations, including gauge fluctuations,
separately for each phase. In particular, because of the low dimensionality, and since the 
energy scales of all 
the terms in the renormalized Hamiltonian scale with $J$ and are comparable, the
mean-field approximation would not work near the transitions. For this reason,
probing the nature of the transition near $T^*$, and the
calculation of $T_c(x)$ or, possibility of a Nernst effect, would require a more sophisticated 
treatment. We also do not discuss the issue of collective modes, the treatment of which
would require going beyond the MF approximations. There are a number of
such collective excitations. An important one is the physical electron itself, which can
be probed by the photoemission spectrum. Experimentally one sees something like a Fermi
surface above $T^*$, and a Fermi arc in the pseudogap normal state, although there are
no quasiparticle excitations associated with these. One important issue is the possible
appearance of sharp peaks in the gap region of the superconducting state. Our theory 
does not rule out the emergence of a more conventional Fermi liquid state at large 
$x$, where the physical electron would appear as a collective excitation via 
spin-charge recombination \cite{sar3}. These and other issues would be discussed 
in a future paper.

\end{document}